\documentclass[preprint2, trackchanges]{aastex631}
\usepackage{placeins}
\usepackage{ltablex}
\keepXColumns

\usepackage{longtable}
\usepackage{rotating}
\usepackage{graphicx}
\usepackage{booktabs}
\usepackage{amsmath}
\usepackage{textcomp}
\usepackage{booktabs}
\usepackage{array}
\begin{document}

\hypersetup{linkcolor=blue,citecolor=blue,filecolor=cyan,urlcolor=blue}

\newcommand{\vdag}{(v)^\dagger}
\newcommand\aastex{AAS\TeX}
\newcommand\latex{La\TeX}


\title{Multi-Frequency Study of FRB20201124A with the uGMRT }
\author[0009-0005-4130-892X]{C. Dudeja}
\affiliation{National Centre for Radio Astrophysics, Pune (411007), Maharashtra, India}

\author[0000-0002-2892-8025]{J. Roy}
\affiliation{National Centre for Radio Astrophysics, Pune (411007), Maharashtra, India}

\author[0000-0002-2441-4174]{U. Panda}
\affiliation{National Centre for Radio Astrophysics, Pune (411007), Maharashtra, India}

\author[0000-0003-0669-873X]{S. Bhattacharyya}
\affiliation{National Centre for Radio Astrophysics, Pune (411007), Maharashtra, India}
\affiliation{Department of Computer Science and Engineering, SR University, Warangal, Telangana 506371, India}
\correspondingauthor{Chahat Dudeja}
\email{cdudeja@ncra.tifr.res.in}

\begin{abstract}

We present results from a monitoring campaign of FRB 20201124A using the uGMRT during its active phase between MJD 59342 and 59362. Simultaneous observations in Band 4 (550--950 MHz) and Band 5 (1060--1460 MHz) provide a uniform view of the source activity. The burst activity exhibits a frequency-dependent evolution, with the last Band-5 burst detected on MJD 59358, while Band-4 activity persisted until MJD 59362. We identified burst pairs in the simultaneously observed Band-4 and Band-5 data, with time separations of 31.8 ms and 52.3 ms. These separations are comparable to the short-timescale component of the waiting-time distribution, which peaks at $\sim$ 60 ms, suggesting that the multi-band burst pairs may be associated with the same short-timescale clustering process. Contemporaneous FAST observations detected the Band-5 bursts but revealed no strictly simultaneous counterparts to the corresponding Band-4 bursts. These results suggest frequency-dependent burst occurrence and are consistent with patchy spectral occupancy over the 550--1460 MHz frequency range. The detected bursts exhibit diverse temporal and spectral morphologies. Consecutive bursts were observed with separations of 16.7(2)--291.5(6) ms, revealing short repetition intervals and possible sub-second quasi-periodicity. The waiting-time and energy distributions are bimodal, indicating at least two characteristic emission timescales and energy modes. Burst fluences span 1.7(3)--78(3) Jy ms, and the cumulative fluence distribution is well described by a broken power law. These results provide a simultaneous multi-frequency view of FRB 20201124A during its May 2021 active episode and support a picture in which the source exhibits complex frequency-dependent activity and broadband emission behaviour.

\end{abstract}

\section{Introduction} 
Fast Radio Bursts (FRBs) are transient radio pulses of extragalactic origin, lasting only a few milliseconds, with isotropic-equivalent energies typically in the range of  $10^{37}$ to $10^{42}$ erg \citep{Petroff_2019, Zhang2020}. Their dispersion measures (DMs) generally exceed the expectations from Galactic models. \citep{Lorimer2007, Petroff_2019, Xiao, cordes}. FRBs are classified into one-off and repeating sources; however, recent studies suggest that some apparently non-repeating events may have very low repetition rates \citep{yamasaki2023true, Kirsten2024, 2026MNRAS.545f1937O}. The physical origin of FRBs is still unknown, with proposed models including magnetars, compact object mergers, and binary interactions \citep{Platts2019, Zhang2020, Nimmo2022,Bochenek2020, 2010vaoa.conf..129P}. Statistical properties such as burst energy and waiting times (time difference between adjacent bursts) have been used to test these models \citep{Zhang_2021, Oppermann2018, cruces2020}.
\\
FRB 20201124A was discovered by CHIME/FRB in November 2020 and showed a large number of bursts with diverse temporal and spectral structures \citep{Lanman_2022, Xu2022}. It exhibits variable polarization, including circular polarization up to 90\% \citep{Niu_2024, Jiang2024,  Hilmarsson_2021}, and is associated with a persistent radio source \citep{PRS}. These features are consistent with emission from a highly magnetized environment, such as a magnetar magnetosphere. \citet{chen_du2024} reported a relatively high surface magnetic field strength, and suggested the presence of multiple emission regions within the magnetosphere. The source also shows significant temporal variations in its rotation measure (RM), indicating a dynamic magneto-ionic environment \citep{Xu2022}. \citet{2025arXiv250517880Z} identified two distinct epochs (MJD
59310 and MJD 59347) where FRB 20201124A exhibited second-level periodicity in its burst arrival times that coincides with critical RM evolution, which may reflect changes in magnetic field topology or line-of-sight geometry. Using a larger sample of bursts, \citet{Niu2022} reported no evidence for periodicity, while \citet{2025arXiv250514219G} also found no statistically significant periodic signal. A magnetar–Be star binary system has been proposed as a viable model to explain these features \citep{Wang2022}.
\\
Several repeating FRBs have shown burst energy distributions that follow power law or log normal forms \citep{Li2021, Aggarwal_2021, Qin}. Waiting times between adjacent bursts from repeating FRB sources are often inconsistent with a Poisson process and instead exhibit clustered or power-law-like distributions, suggestive of burst storms or active windows \citep{metzger, cruces2020, Oostrum_2020, Xu2022}. Similar behaviour has been observed in repeaters such as FRB 121102 and FRB 180916.J0158+65, where burst clustering and structured energy distributions have been interpreted within magnetar based models \citep{Li2021, Nimmo2022}.
\\
The waiting time distribution of FRB 20201124A appears bimodal, suggesting the presence of two characteristic burst timescales corresponding to closely spaced bursts and longer quiescent intervals \citep{Zhang_2022, Xu2022}. Bimodality in burst energy or waiting time distributions may indicate the presence of distinct emission mechanisms, such as magnetospheric versus shock driven processes, or could result from modulation effects in a binary system \citep{Li2021, 2025arXiv250216626L}.
\\
Previous studies of FRB 20201124A have combined observations from multiple facilities with different sensitivities and frequency coverage \citep{Xu2022, Lanman_2022, Kirsten2024, Hilmarsson_2021, Zhang_2022}. In contrast, this work focuses on a uniform simultaneous multi-frequency dataset obtained with a single instrument. 
Section 2 describes the observational setup and data collection. Section 3 explains the data analysis techniques used. Section 4 presents the main results, while Section 5 compares these findings with those from other FRBs. Section 6 explores possible physical interpretations, and Section 7 summarises the key conclusions of this work.

\begin{deluxetable*}{clcccc}
\tablecaption{GMRT Observations Summary
\label{tab:gmrt_summary}}
\tablehead{
\colhead{Date (2021) (MJD)} & 
\colhead{Beam Mode} & 
\colhead{Obs. Dur. (h)} & 
\colhead{Freq. Channels} & 
\colhead{Freq. Range (MHz)} & 
\colhead{Time Res. ($\mu$s)}
}
\startdata
 8 May (59342.349309)  & PA       & 1.3   & 8192               & 550–950,\ 1060–1460                    & 163.84,\ 163.84 \\
 9 May (59343.281212)  & PA       & 1.3   & 8192               & 550–950,\ 1060–1460                    & 163.84,\ 163.84 \\
18 May (59352.271998) & PA, IA   & 1.0   & 8192               & 550–950,\ 1060–1460                    & 327.68,\ 327.68 \\
24 May (59358.299619) & PA, IA   & 1.4   & 8192               & 550–950,\ 1060–1460                    & 327.68,\ 327.68 \\
28 May (59362.248791) & PA, IA   & 2.0   & 8192               & 550–950,\ 1060–1460                    & 327.68,\ 327.68 \\
12 June (59377.356705) & PA   & 4.0   & 4096          & 550–750                                & 81.92 \\
14 June (59379.394803)& PA, CD   & 4.4   & 4096,\ 2048         & 300–500,\ 550–750,\ 1060–1260         & 327.68,\ 40.96 \\
15 June (59380.247709)& PA, CD   & 3.23  & 4096,\ 2048         & 300–500,\ 550–750,\ 1060–1260         & 327.68,\ 40.96 \\
\enddata

\tablecomments{Typical fluence detection thresholds for the Band-4 PA, IA, and CD observations were approximately 0.013, 0.055, and 0.013 Jy ms, respectively, estimated using the radiometer equation, assuming a detection threshold of S/N = 6.5 and a representative burst width of 1 ms.}

\end{deluxetable*}

\begin{figure*}[ht!]
\centering
\includegraphics[width=\textwidth]{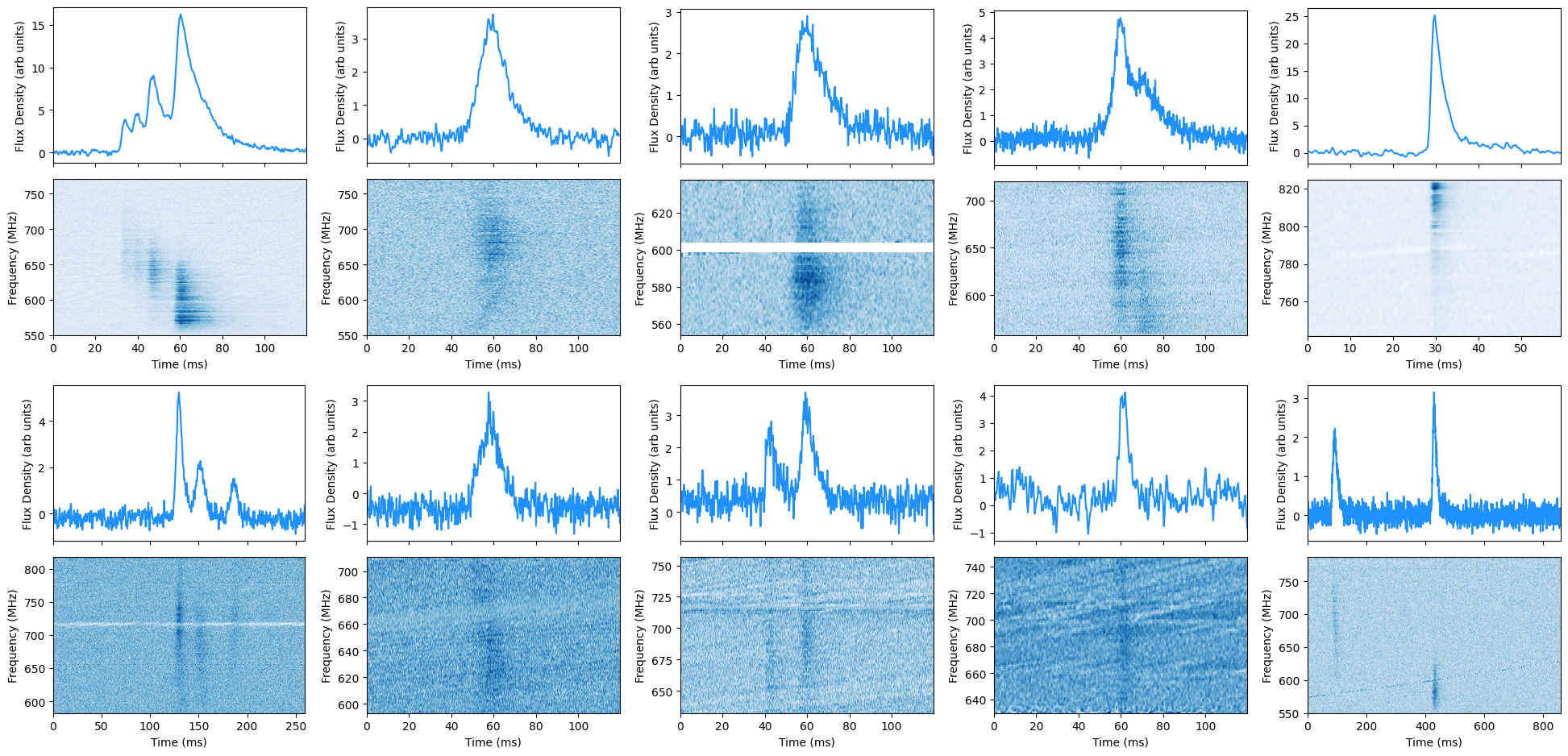}
\caption{A gallery of bursts detected in our observations, displaying dynamic spectra (lower panels) and their dedispersed time profiles (upper panels). All bursts were detected in uGMRT Band-4 observations (550--950 MHz), and the displayed frequency ranges have been cropped to highlight the burst emission.} The examples highlight a wide range of temporal and spectral structures, including narrowband and broadband emission, multi-component, sub-components and closely spaced independent components, faint and bright bursts, short- and long-duration events, highly scattered and less scattered profiles, as well as frequency-drifting features.
\label{fig:1}
\end{figure*}

\begin{figure*}[ht!]
\centering
\includegraphics[width=\textwidth]{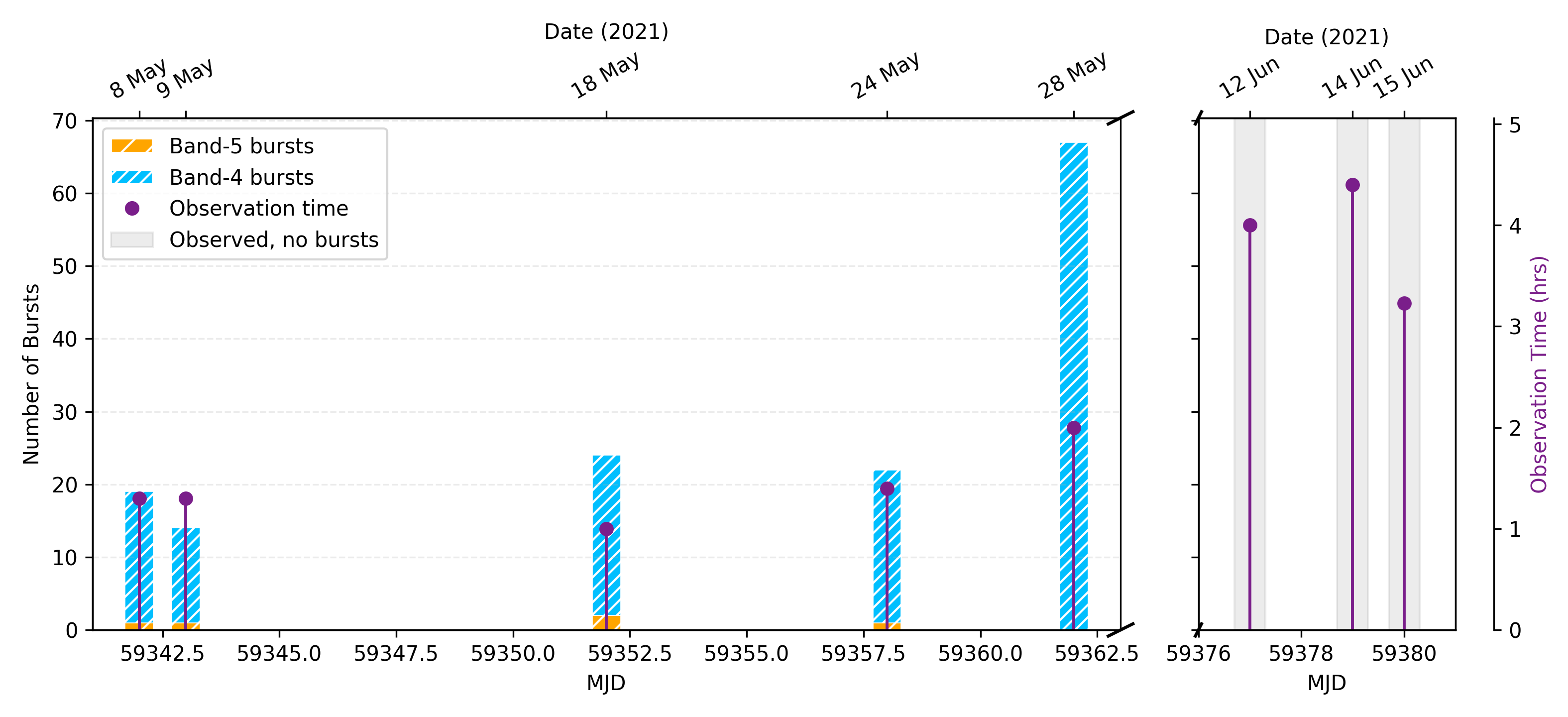}
\caption{Burst activity of FRB~20201124A during May–June 2021. The x-axis shows Modified Julian Dates (MJDs) and is broken between MJD 59362 and 59377 to indicate the 15-day gap between observing sessions. The top x-axis shows the corresponding calendar dates. The bar plots represent the number of detected bursts per observing session, with Band-5 (orange, hatched) at the base and Band-4 (blue, hatched) stacked above. The purple markers (right y-axis) indicate the total on-source time for each session, obtained in sub-array mode (simultaneous Band-4 and Band-5 observations). The grey shaded regions correspond to the June observations (12, 14, and 15 June 2021), during which no bursts were detected despite a total on-source time of $\sim$11.6 hours. In contrast, a total of 146 bursts (141 in Band-4 and 5 in Band-5) were detected over $\sim$7 hours of observations during May (MJD 59342–59362).}
\label{fig:widefig}
\end{figure*}

\section{Observations}

We observed FRB~20201124A using the uGMRT under the DDT observation DDTC171 and DDTC182 over eight sessions between 8~May and 15~June 2021, with a total of approximately 18.6 hours of on-sky time. The source position was taken to be at RA = 05$^{\mathrm{h}}$\,08$^{\mathrm{m}}$\,03.5074$^{\mathrm{s}}\pm 2.7$mas, Dec = +26$^\circ$\,03$'$\,38.5052$''\pm 2.6$mas (J2000) \citep{Nimmo_2022}. Observations were carried out using a combination of beamforming modes, including Phased Array (PA), Incoherent Array (IA), and Coherent Dedispersion (CD), to investigate the burst properties across a range of temporal and spectral resolutions.
\\
During the first set of observations (8–28~May 2021), data were recorded in PA and IA modes with 8192 frequency channels. A simultaneous sub-array configuration was used to observe Band~4 (550–950~MHz) and Band~5 (1060–1460~MHz), each with a bandwidth of 400~MHz. 
The time resolution during this period ranged from 163.84~$\mu$s to 327.68~$\mu$s, sufficient to resolve millisecond-scale burst features.
\\
In June, the observing configuration was modified to include CD mode for higher time resolution and to capture finer temporal structure in the bursts. These observations spanned Band~3 (300–500~MHz), Band~4 (550–750~MHz), and Band~5 (1060–1260~MHz), using 4096 and 2048 frequency channels, as listed in Table~1. Each band was recorded with a bandwidth of 200~MHz. The time resolution in CD mode was brought down to 10.24~$\mu$s, allowing a more detailed study of short-duration burst components. For the CD observations, the data were coherently dedispersed in real time using a reference DM of 414.0~pc~cm$^{-3}$.
\\
For flux calibration, a test pulsar,  J0543$+$2329, located close to the target region, was observed for 5 minutes prior to each FRB scan and used to estimate the gain of the PA beam using radiometer equation \citep{lorimer_handbook}.

\section{Data Analysis}
We initially processed the data collected on May 8, 2021 and mitigated the radio frequency interference (RFI), using the GMRT Pulsar Tool (\href{https://github.com/chowdhuryaditya/gptool} {\texttt{gptool}}\footnote{\url{https://github.com/chowdhuryaditya/gptool}}). The raw data were converted into SIGPROC\footnote{{\url{https://github.com/SixByNine/sigproc}}} filterbank (`.fil') files, followed by incoherent dedispersion using PRESTO's \href{https://github.com/scottransom/presto/blob/master/src/prepsubband.c}{\texttt{prepsubband}}\footnote{\label{presto-footnote}\url{https://github.com/scottransom/presto}} utility. This process covered a DM range of 400 to 430 pc cm$^{-3}$ with 1 pc cm$^{-3}$ steps. We imposed constraints on signal width ($\leq$65.536 ms) and signal-to-noise ratio (SNR $\geq$6.5$\sigma$). To expedite processing, the data were downsampled from a time resolution of 163.84~$\mu$s or 327.68~$\mu$s to 655.36~$\mu$s. Single-pulse searches were conducted using PRESTO's \href{https://github.com/scottransom/presto/blob/master/bin/single_pulse_search.py}{\texttt{single\_pulse\_search.py}}\footref{presto-footnote} tool.
\\
We then extracted features such as frequency-integrated profiles, dynamic spectra (waterfall plots), and DM-time plots using \href{https://github.com/thepetabyteproject/your}{\texttt{your} \footnote{\label{your}\url{https://github.com/thepetabyteproject/your}}}. FRB pulses were manually identified by examining band-limited patterns in dynamic spectra, signal-like peaks in frequency-integrated profiles, and bow-tie structures in DM-time plots. A few bursts are shown in Figure ~\ref{fig:1}, while the remaining bursts can be found in the appendix section.
\\
This procedure was applied to all observation epochs. For data from 8 and 9 May 2021 (initial time resolution of 163.84 $\mu$s), we downsampled it to 655.36 $\mu$s. Similarly, data from May 18, 24, and 28, 2021 (initial time resolution of 327.68 $\mu$s) were also downsampled to 655.36 $\mu$s. On these dates, data from the IA beam were also recorded. We first searched for bursts in the PA mode and subsequently formed a Post-Correlation (PC) beam \citep{2018ApJ...864..160R} by subtracting the IA beam from the PA beam. Given its better interference immunity due to the removal of autocorrelation terms, the PC beam revealed additional bursts that were not visible in the PA data.
\\
Data from June 12, 14, and 15, 2021, with initial time resolutions of 81.92 $\mu$s and 40.96 $\mu$s, were similarly downsampled to search for bursts.
\\
In total, we detected 141 bursts in Band-4 (550–950 MHz) and 5 bursts in Band-5 (1060–1460 MHz) from the May epochs. No bursts were detected on June 12, 14, and 15 in the 300-1460 MHz range, indicating a cessation of activity. At the time of initial classification, 133 bursts were identified in Band-4; however, during subsequent profile modeling, we identified additional nearby bursts. Bursts separated by off-pulse noise were identified as an independent events, whereas closely spaced structures without intervening off-pulse noise were treated as sub-bursts of the same event. This increased the number of Band 4 bursts to 141, as shown in Figure ~\ref{fig:widefig}. 
\\
Using a detection threshold of S/N $\geq$ 6.5$\sigma$, an analysis of the Band-4 bursts revealed that 11.2\% had SNR \textbf{values in the range 6.5–}10$\sigma$, 32.8\% were in the range of 10$\sigma$–15$\sigma$, 15\% were in the range of 15$\sigma$–20$\sigma$, and the remaining 41\% exhibited an SNR $\geq$ 20$\sigma$. Notably, the source ceased activity after May 28, consistent with reports from the FAST \citep{Xu2022}, which detected no bursts between May 28 and June 11.

\subsection{RFI Mitigation and DM Refinement}
Before refining the DM of each burst, we applied a Radio Frequency Interference (RFI) mitigation step to clean the data. This step was important to ensure that RFI do not affect the accuracy of DM estimation or burst modelling. We used a method that combines spectral kurtosis and Savitzky-Golay filtering, following the approach described in \href{https://github.com/thepetabyteproject/your/blob/main/bin/your_rfimask.py}{\texttt{your}\footref{your}}. Spectral kurtosis identifies and masks RFI-contaminated frequency channels based on their statistical behaviour, while the Savitzky-Golay filter smooths the dynamic spectrum and enhances the clarity of the bursts. Together, these steps helped remove narrowband and broadband RFI without affecting the actual signal.

\begin{figure}
\centering
\includegraphics[width=\linewidth]{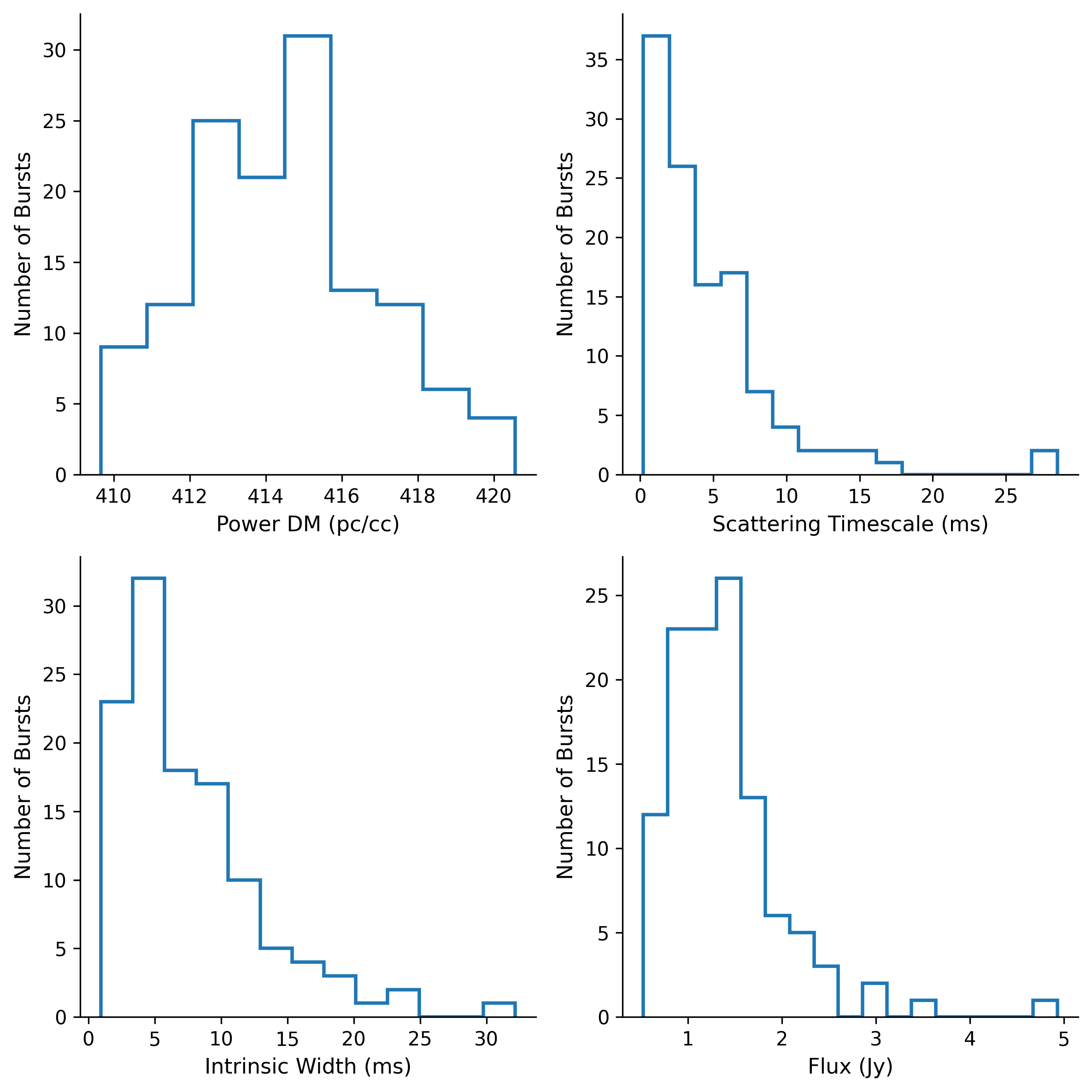}
\caption{Distributions of key burst properties. Top left: DM-Power, showing a concentration around 412–415 pc/cc. Top right: Scattering timescales at a reference frequency of 950 MHz, indicating that most bursts exhibit minimal to moderate scattering. Bottom left: Intrinsic burst durations, with the majority under 10 ms. Bottom right: Burst flux densities, peaking below 2 Jy.}
\label{fig:3}
\end{figure}

Each burst already had an approximate DM value from the  \href{https://github.com/scottransom/presto/blob/master/bin/single_pulse_search.py}{\texttt{single\_pulse\_search.py}}\footref{presto-footnote}. To refine these estimates, we applied the \href{https://github.com/hsiuhsil/DM-power}{DM-Power \footnote{\url{https://github.com/hsiuhsil/DM-power}}} method, which uses the power distribution across different Fourier modes of the burst \citep{2022arXiv220813677L}.  The resulting DM values were selected based on the visual alignment of the burst structures in the dedispersed dynamic spectra and frequency-averaged profiles. We applied the DM-Power method to all bursts, including both simple bursts and bursts with multiple sub-components. The power-based method worked better for bursts with complex shapes, where the signal-to-noise  based approach was less reliable.
\\
After refining the DM, we modelled both the burst profiles and spectra. The burst profile, representing flux density as a function of time, was fitted using a Gaussian function convolved with an exponential tail to account for scattering. For bursts with multiple components, the frequency-integrated profile was modelled using multiple Gaussian components convolved with a common exponential scattering tail. We tested allowing the scattering timescale to vary between components, but found no significant improvement in the fits. Therefore, a single scattering timescale was adopted for all components within a given burst, while the intrinsic widths were fitted independently. The burst spectrum, representing flux density as a function of frequency, was fitted using a Gaussian model to estimate the peak frequency and emission bandwidth. For events with SNR $\geq$ 20 $\sigma$, we used the native time resolutions (163.84 $\mu$s / 327.68 $\mu$s). For events, which had SNR $<$ 20 $\sigma$, we downsampled it to 655.36 $\mu$s and averaged the frequency channels from 8192 to 512 for all the bursts. The distributions of DM-power, intrinsic width, scattering width, and flux are shown in Figure ~\ref{fig:3}. A table containing the fitted burst properties is provided in appendix.
\section{Results}

\subsection{Waiting Time}

\begin{figure}
\centering
\includegraphics[width=\linewidth]{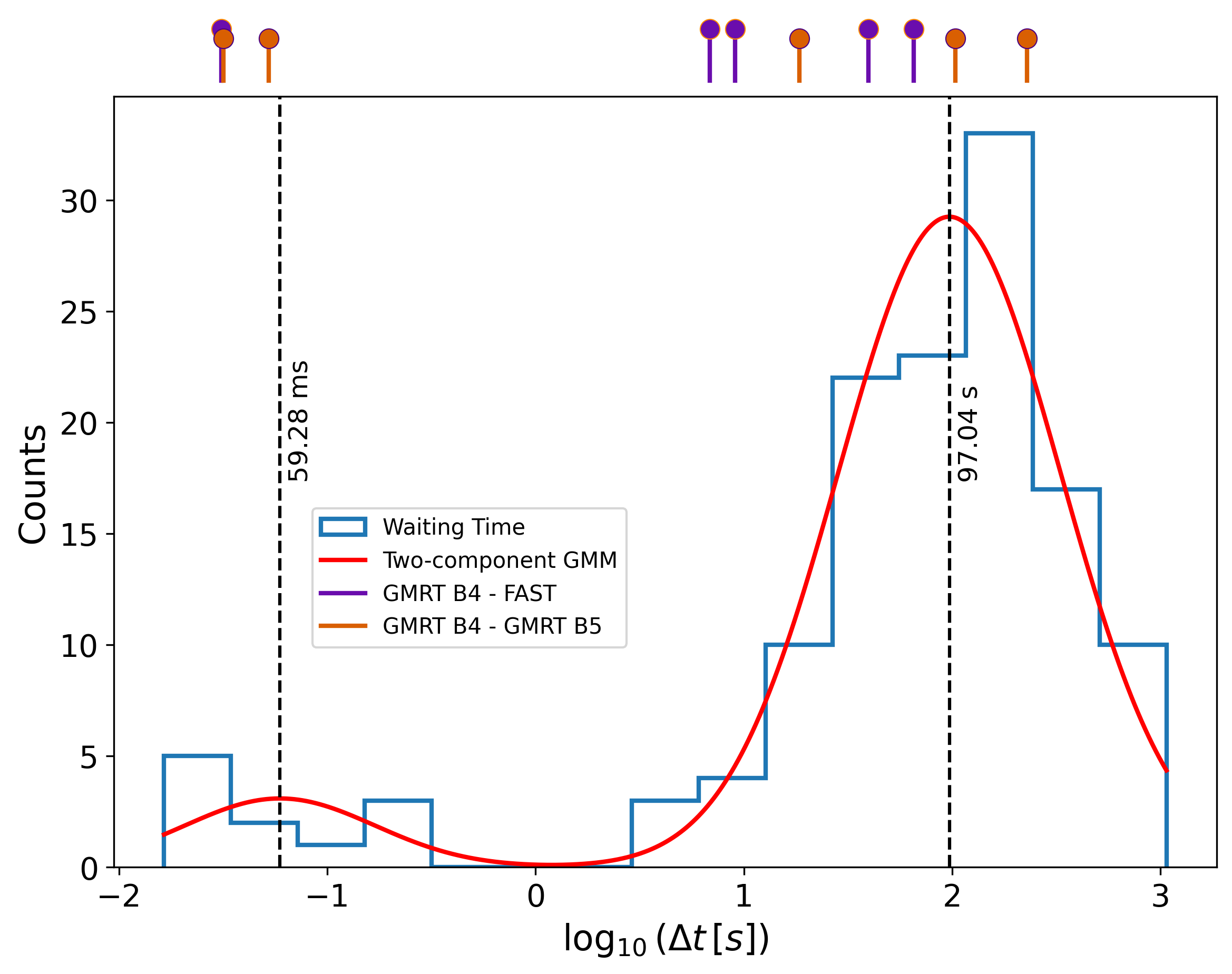}
\caption{Histogram of waiting times (blue), fitted with a two-component  Gaussian Mixture Model (GMM) in log space (red). The black dashed lines correspond to the peak waiting times of the two fitted components at $\sim$ 59 ms and 97 s. Purple markers denote waiting times between burst pairs identified in simultaneous GMRT Band-4 and FAST observations, while orange markers denote waiting times between burst pairs identified in simultaneous GMRT Band-4 and GMRT Band-5 observations. Notably, the GMRT Band-4--Band-5 separations of 31.8\,ms and 52.3\,ms lie within the short-timescale component of the waiting-time distribution.  Vertical lollipop markers above the histogram indicate nearly simultaneous burst detections between GMRT Band-4 and FAST (purple) and between GMRT Band-4 and Band-5 (orange)
}

\label{fig:4}
\end{figure}

Figure~\ref{fig:4} presents the waiting time distribution fitted with a Gaussian Mixture Model (GMM) in log space. The histogram exhibits a clear bimodal structure with two peaks: a short-timescale peak at $\sim$ 60 ms and a long-timescale peak at \textbf{97} sec. The short-timescale peak depends on how sub-bursts are defined. If every identifiable sub-burst peak is counted individually, the distribution shifts to a narrower timescale, peaking at $\simeq$ 23 ms. This represents the minimum separation between fine structures and highlights the rapid variability present in the data. In contrast, if sub-bursts are grouped together and only the arrival time of the strongest peak in each group is used, the distribution broadens and centers around $\simeq$ 60 ms. For the shorter peak, we considered both grouped sub-bursts and closely spaced independent bursts. This approach avoids over-counting blended components and provides a more conservative estimate of short-timescale clustering. The second, much broader peak at $\simeq$ 97 s corresponds to waiting times between fully independent bursts, separated by clear off-pulse regions. The purple and orange markers indicate waiting times derived from burst pairs identified during simultaneous GMRT Band 4--FAST L-Band and GMRT Band 4--GMRT Band 5 observing sessions, respectively. These events span both the short- and long-timescale components of the distribution and are discussed further in Section~4.7. Overall, these results indicate that FRB 20201124A exhibits activity on at least two distinct timescales: tens of milliseconds associated with closely spaced bursts, and hundreds of seconds corresponding to separations between independent bursts. Interestingly, the temporal separations identified between simultaneous Band-4 and Band-5 bursts (31.8 ms and 52.3 ms; Section~4.7) are comparable to the characteristic timescale of the short-timescale component. 

\subsection{Cumulative Burst Rate vs. Fluence}

\begin{figure}
    \centering
    \includegraphics[width=\linewidth]{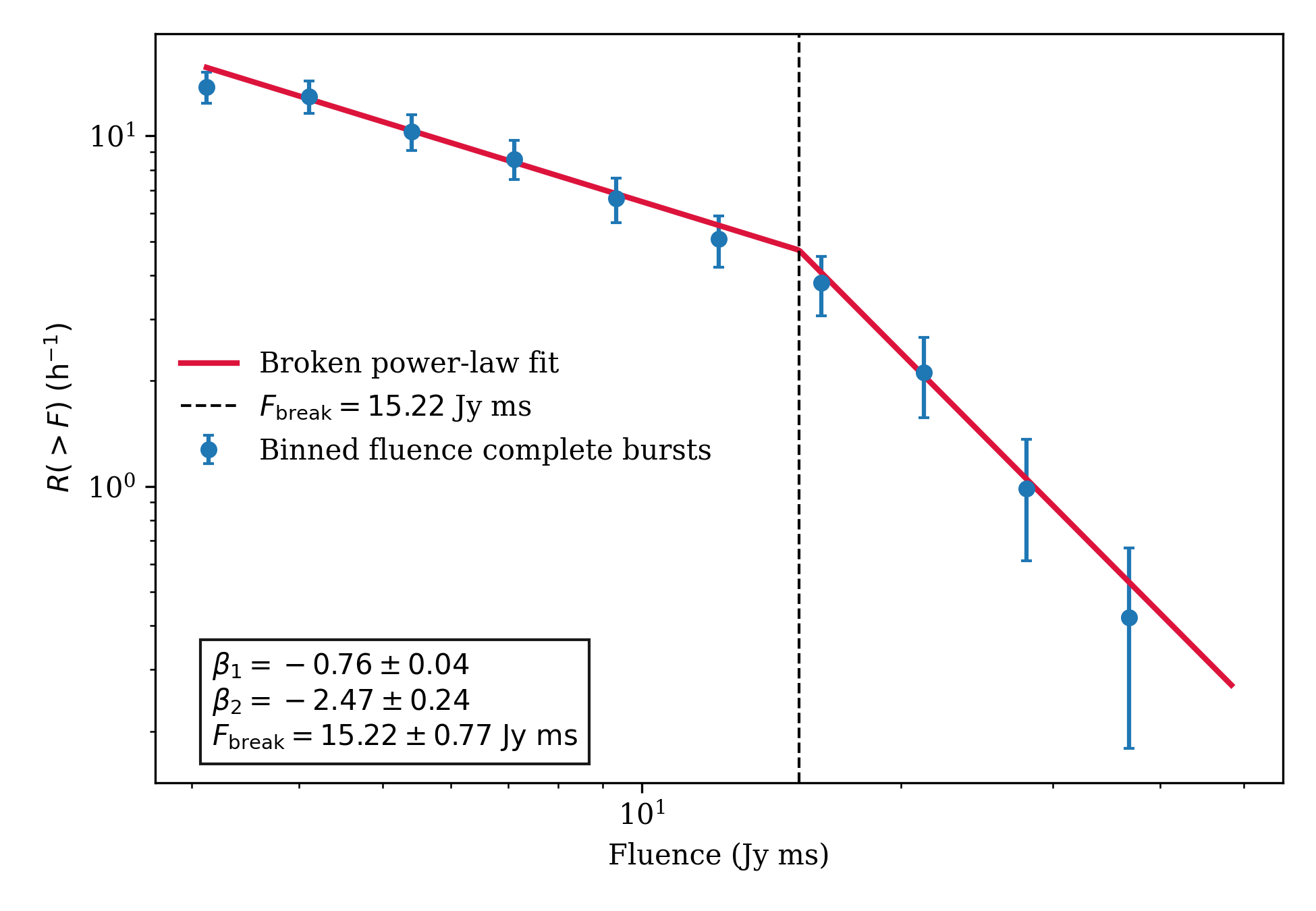}
    \caption{Cumulative burst rate above a fluence threshold for fluence-complete bursts (fluence $>3.05$~Jy\,ms), plotted in log-log scale. A broken power-law fit to the cumulative distribution shows a slope change at $F_\mathrm{break} = 15.22 \pm 0.77$~Jy\,ms. The best-fitting differential power-law indices are $\alpha_1 = -1.76 \pm 0.04$ and $\alpha_2 = -3.47 \pm 0.24$, corresponding to cumulative slopes of ${-0.76 \pm 0.04}$ and ${-2.47 \pm 0.24}$.}
    \label{fig:5}
\end{figure}

Figure~\ref{fig:5} shows the cumulative burst rate as a function of fluence for fluence-complete bursts. To correct for observational biases, we applied a fluence completeness threshold following the method described in \citet{10.1093/mnras/stu2650} for S/N threshold of 6.5 $\sigma$ and the maximum observed pulse width of 33.73 ms . Only 109 bursts in Band-4 with fluence values above 3.05~Jy\,ms were included to ensure statistical completeness. These complete fluence values were sorted and binned logarithmically into ten intervals. The burst rate in each bin was computed by dividing the number of bursts by the total observing time, which summed to 7.1 hours.
\\
To describe the burst fluence distribution, we fitted the cumulative distribution using a broken power-law model of the form
\[
R(>F) \propto
\begin{cases}
F^{(\alpha_1 + 1)}, & F < F_\mathrm{break} \\
F^{(\alpha_2 + 1)}, & F \geq F_\mathrm{break}
\end{cases}
\]
where \( F \) is the fluence, \( F_\mathrm{break} \) is the fluence at which the slope changes, and \( \alpha_1, \alpha_2 \) are the fitted cumulative power-law indices. We found the break to occur at \( F_\mathrm{break} = 15.22 \pm 0.77~\mathrm{Jy\,ms} \), with best-fitting differential indices \( \alpha_1 = -1.76 \pm 0.04 \) and \( \alpha_2 = -3.47 \pm 0.24 \). This corresponds to \textbf{cumulative} slopes \({-0.76 \pm 0.04} \) below the break and \({-2.47 \pm 0.24} \) above the break.
\\
The sharp change in slope suggests a transition in the underlying burst-generation mechanism. The shallower slope at lower fluences indicates that faint bursts are more common, while the steep drop-off at higher fluences implies that bright bursts are increasingly rare. Such a broken power-law behavior has been observed in other repeating FRBs, pointing to potential similarities in their emission physics.

\subsection{Energy Distribution}

The isotropic energy for each burst is calculated using the following formula:

\[
E = \frac{4 \pi D_L^2 F \Delta \nu}{1 + z}
\]
where \( F \) is the fluence, defined as the product of flux and equivalent width. The flux is measured using the radiometer equation. Equivalent width is the quadrature sum of burst width and scattering timescales. \( D_L \) is the luminosity distance, \( \Delta \nu \) is the emission bandwidth, and \( z \) is the redshift of the source.
The luminosity distance for the source was taken to be \( 453.3 \, \text{Mpc} \), and the redshift is \( z = 0.09795 \) \citep{2021ApJ...919L..23F, 2021A&A...656L..15P, Nimmo_2022, 10.1093/mnras/stac465}. Since due to RFI conditions, we could measure the emission bandwidth only for 50\% of the bursts, for the calculation of energy, we have taken the usable bandwidth, which is \( 200 \, \text{MHz} \) instead of emission bandwidth.


\begin{figure}
\centering
\includegraphics[width=\linewidth]{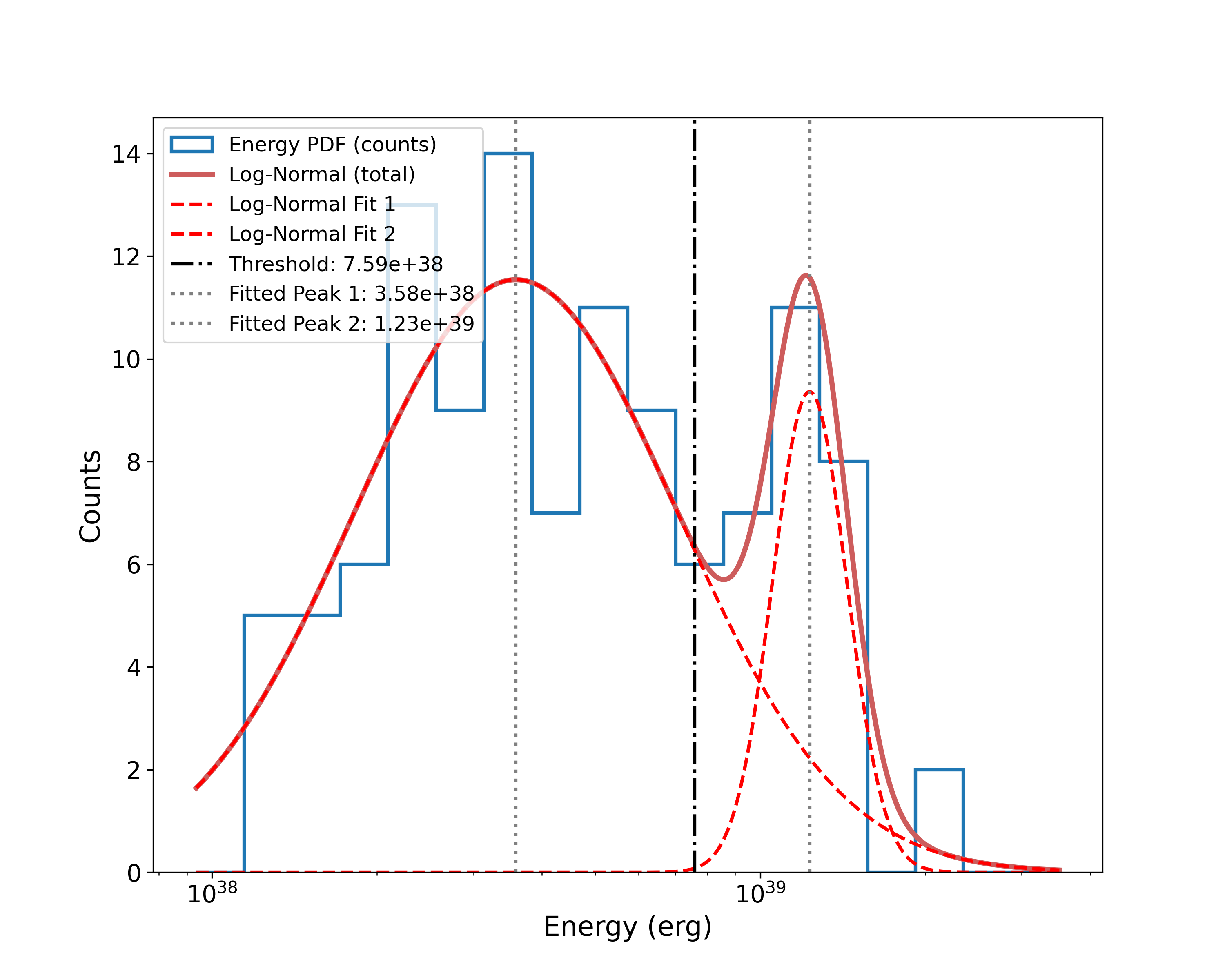}
\caption{Distribution of burst energies on a logarithmic scale. The histogram shows a bimodal structure, with two log-normal fits indicating two distinct peaks at \( 3.58 \times 10^{38} \,\mathrm{erg} \) and \( 1.23 \times 10^{39} \,\mathrm{erg} \). The black dash-dotted line corresponds to an energy threshold of \( 7.59 \times 10^{38} \,\mathrm{erg} \), taken from the break in the fluence distribution (Figure~\ref{fig:5}). }
\label{fig:6}
\end{figure}

The distribution of burst energies, shown in Figure~\ref{fig:6}, displays a clear bimodal structure when plotted on a logarithmic scale. A log-normal fit identifies two distinct peaks, with the first centred at \( 3.58 \times 10^{38} \,\mathrm{erg} \) and the second at \( 1.23 \times 10^{39} \,\mathrm{erg} \), suggestive of two underlying populations. The black dash-dotted line, corresponding to the break in the fluence distribution (Figure~\ref{fig:5}), corresponds to an energy threshold of \( 7.59 \times 10^{38} \,\mathrm{erg} \). This threshold lies between the two fitted peaks and intersects the observed counts, although it does not align with the exact minimum of the fitted log-normal components. This behaviour may point to a change in burst properties, separating more frequent, lower-energy events from rarer, higher-energy ones. Since energy and fluence are directly related for a fixed fiducial bandwidth, the broken power-law in the fluence distribution and the appearance of two peaks in the energy distribution are closely related and should not be considered independently. Together, however, these properties suggest the presence of distinct burst-generation mechanisms.

\begin{figure}
\centering
\includegraphics[width=\linewidth]{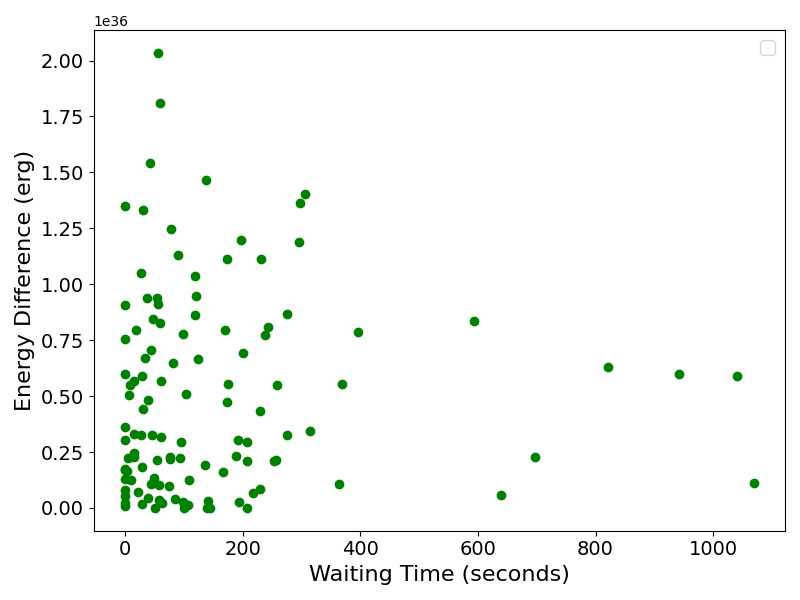}
\caption{Difference in isotropic-equivalent burst energy as a function of waiting time. No significant correlation is observed, with a Pearson correlation coefficient of 0.0189 and a p-value of 0.84.}
\label{fig:7}
\end{figure}

Figure~\ref{fig:7} shows the relationship between the difference in isotropic-equivalent burst energy and the waiting time for FRB 20201124A. Since energy values were not available for all bursts, only those pairs where both adjacent bursts had valid energy estimates were considered. The corresponding waiting time was computed as the time difference between these adjacent bursts. The plot does not exhibit any clear trend or structure, indicating a lack of correlation between the two quantities. A Pearson correlation coefficient of 0.0189 with a p-value of 0.84 confirms the absence of a statistically significant relationship. The absence of a statistically significant correlation between burst energy and waiting time suggests that longer waiting times do not necessarily lead to more energetic bursts. However, given the limited sample size and the complexity of the underlying emission process, stronger conclusions cannot be drawn.

\subsection{Frequency Dependent Activity Cessation}

\begin{figure}
    \centering
    \includegraphics[width=\linewidth]{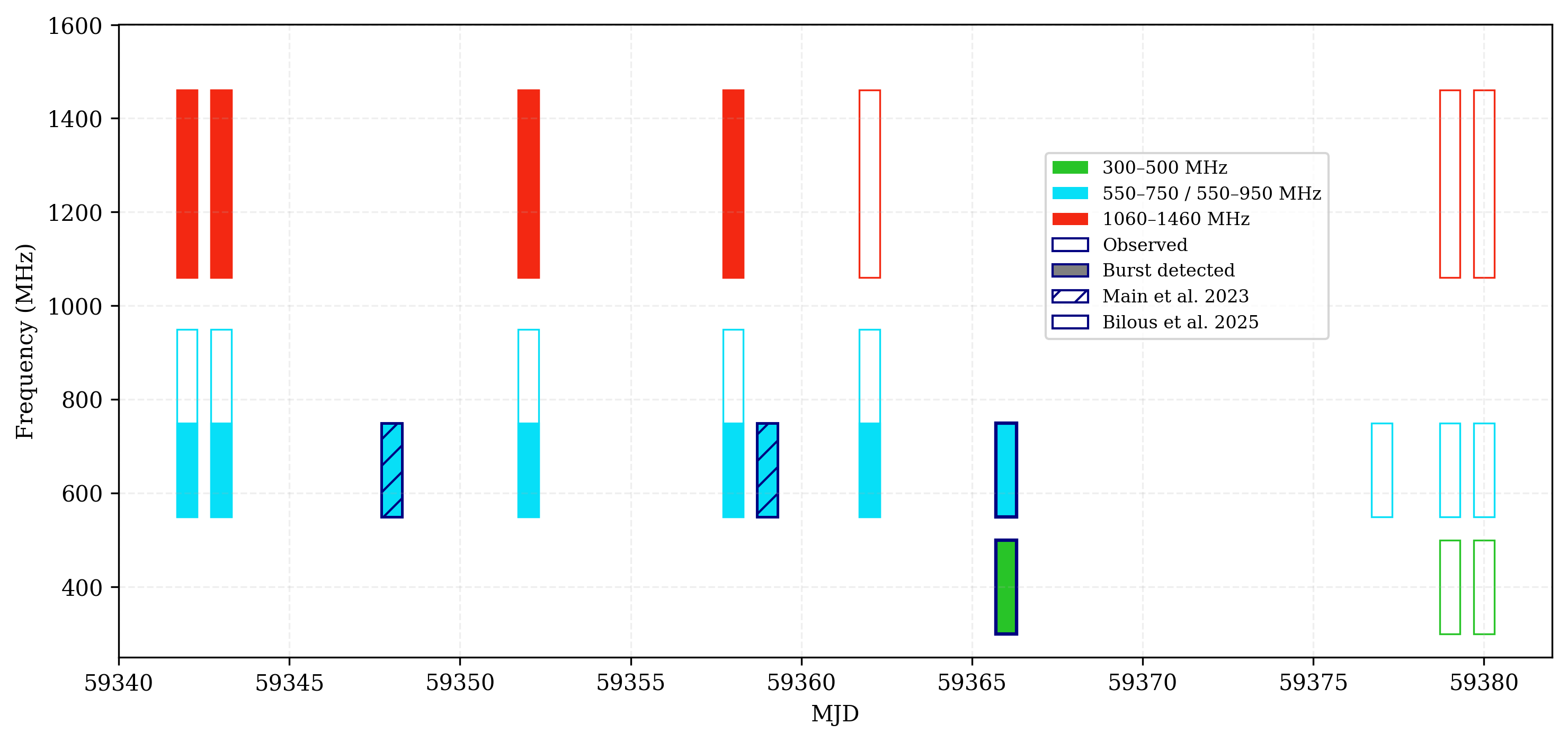}
    \caption{Frequency coverage and burst detections of FRB 20201124A during May--June 2021. Open blocks indicate the full frequency ranges covered by the observations, while filled blocks indicate the frequency ranges over which bursts were detected. The colors denote the observing bands: 300--500 MHz (green), 550--750 / 550--950 MHz (blue), and 1060--1460 MHz (red). Hatched blocks indicate observations from \cite{10.1093/mnrasl/slad036}, and outlined blocks indicate observations from \cite{2025A&A...696A.194B}; unhatched blocks correspond to this work. In particular, the open 550--950 MHz blocks represent the full Band-4 observing bandwidth, whereas the filled 550--750 MHz blocks indicate the frequency range containing detected bursts.}
    \label{fig 8}
\end{figure}

Figure~\ref{fig 8} shows both the frequency coverage and burst detections of FRB20201124A during the active phase observed between 8 May and 4 July across different uGMRT bands. The upper panel shows the observing epochs and frequency ranges covered by uGMRT Band-3 (300--500 MHz), Band-4 (550--750 MHz and 550--950 MHz), and Band-5 (1060--1460 MHz). Solid blocks denote observations from this work, while hatched and outlined blocks indicate observations reported by \citet{10.1093/mnrasl/slad036} and \citet{2025A&A...696A.194B}, respectively. The lower panel shows the corresponding burst detections at the same epochs and frequency ranges. Presenting the observing coverage and detections together allows a direct distinction between epochs with no detected bursts and epochs with no observations.
\\
Although Band-5 was observed only on a few epochs, it revealed an important trend. The last Band-5 burst occurred on 24 May, and no further high-frequency activity was seen afterwards. FAST, which also operates in the L-band (1000-1500 MHz), reported its final detection on 26 May \citep{Xu2022}, consistent with the picture that high-frequency emission ceased around this time. In contrast, Band-4 remained active for longer. On 28 May, a large number of bursts were detected in Band-4 despite the absence of contemporaneous Band-5 detections, and additional Band-3 and 4 bursts were detected on 1 June \citep{2025A&A...696A.194B}. After this epoch, no further bursts were detected despite continued observational coverage at several frequencies.
\\
Taken together, these results independently support the chromatic activity cycle previously noted by \citet{Kirsten2024}, and \citet{2025A&A...696A.194B}, wherein higher-frequency emission ceased earlier than lower-frequency emission. Our uGMRT multi-band observations therefore strengthen the evidence for a frequency-dependent evolution of the source’s activity as it entered the quiescent phase.

\subsection{Short-Timescale Repetition}

\begin{figure}
    \centering
    \includegraphics[width=\linewidth]{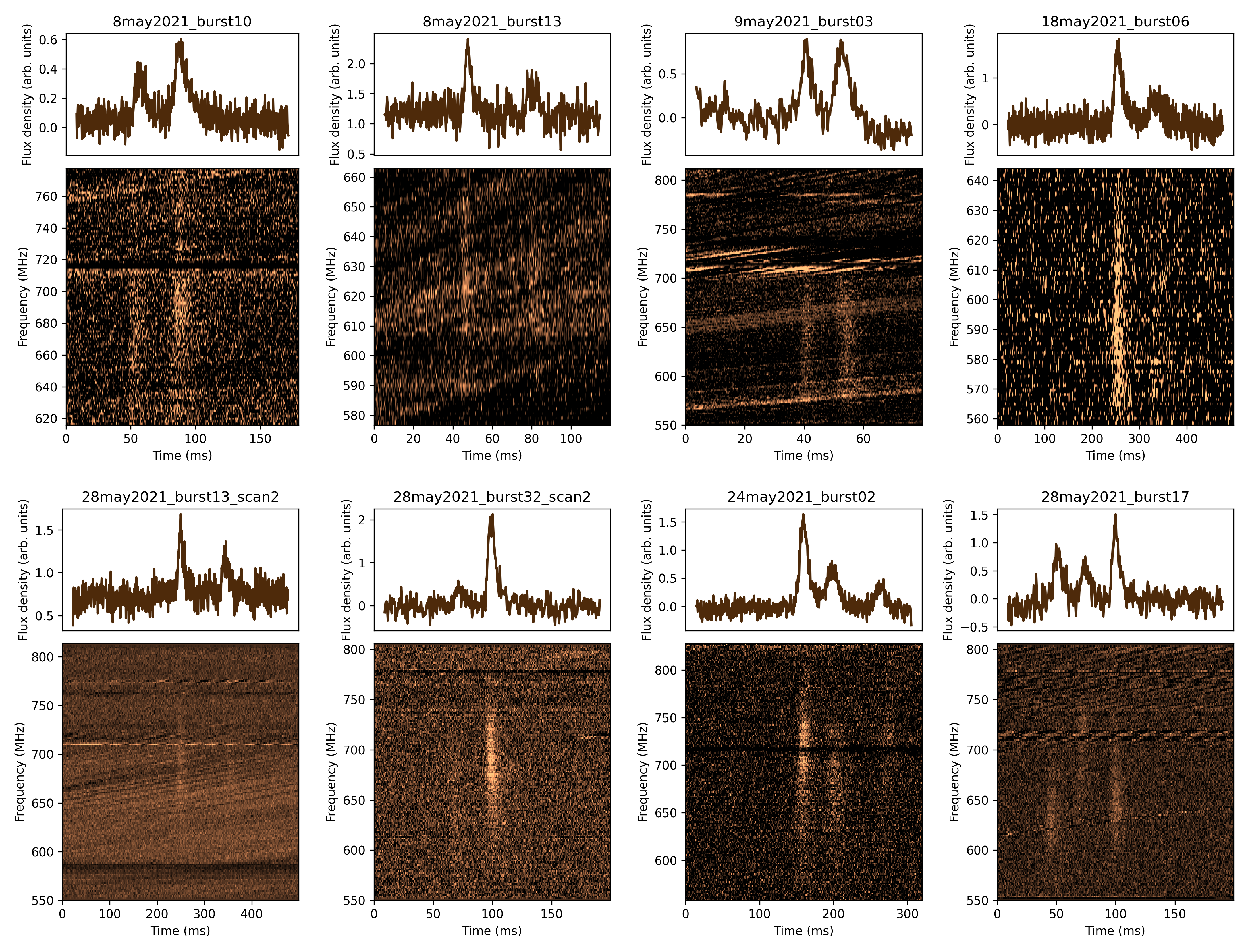}
    \caption{Representative examples of consecutive, independent bursts with arrival-time separations of a few tens to a few hundred milliseconds. Each panel shows the dedispersed time profile (top) and dynamic spectrum (bottom). The bursts are separated by clear off-pulse noise, confirming their independence rather than being sub-components of a single emission episode. }
    \label{fig:9}
\end{figure}

In addition to the diverse burst morphologies presented in Figure~\ref{fig:1}, we identified several instances of consecutive bursts from FRB 20201124A occurring within sub-second intervals (Figure~\ref{fig:9}). These events exhibit arrival-time separations of a few tens to a few hundred milliseconds and are clearly separated by off-pulse noise, confirming that they represent independent bursts rather than components of a single emission feature. A wider separation of $\sim$290 ms is also shown in the burst gallery (Figure~\ref{fig:1}). Such detections demonstrate that the source can generate multiple bursts in rapid succession. 
\\
The detection of these closely spaced bursts further characterizes the temporal behaviour of FRB 20201124A. Similar short-timescale burst separations have previously been discussed by \citet{Niu2022} and \citet{2025A&A...696A.194B}, who noted that such events are compatible with a range of possible repetition intervals. While a quasi-periodicity of $\sim$1.7 s has been suggested based on FAST observations \citep{chen_du2024}, the interpretation of this periodicity remains under discussion. Our observations show that the source is not confined to a single repetition interval. Instead, the occurrence of independent bursts separated by only tens to hundreds of milliseconds raises the possibility that the intrinsic repetition timescale could be shorter, or that the burst engine operates across multiple temporal regimes. Furthermore, \citet{2025arXiv250514219G}, using the same FAST dataset, did not find statistically significant evidence for the reported periodicity, highlighting the sensitivity of the result to the analysis methodology.
\\
Together, the GMRT and FAST results place important constraints on the emission mechanism of FRB 20201124A. The coexistence of sub-second intervals and longer gaps between bursts, as reflected in the bimodal waiting-time distribution, indicates that the source does not follow a uniform periodic clock. Instead, the emission behaviour is best explained by multiple characteristic timescales or a stochastic underlying process.

\subsection{Periodicity Search}

We conducted an autocorrelation function (ACF) analysis on the burst arrival times to search for potential periodicity, following the method described by \citet{2025ApJ...983L..15P}. To avoid introducing artificial peaks, we excluded ToAs from sub-bursts. The normalised ACF was computed by correlating the time series with itself at various time lags. No prominent peaks were found, indicating the absence of a clear periodic trend in the burst timings.
\\
To further investigate potential periodicity, we performed a Lomb Scargle periodogram (LSP) analysis \citep{1982ApJ...263..835S}, a method well suited for detecting periodic signals in unevenly sampled astronomical time series, which is common in FRB observations. The LSP fits sinusoidal functions across a range of frequencies and measures their statistical significance. We searched for periodic signals over two time intervals: 1 to 1000 milliseconds and 1 to 2 seconds. These ranges were motivated by previous studies suggesting a possible 1.7 second periodicity \citep{chen_du2024,2025arXiv250517880Z}.  No statistically significant periodicity was found in either range, indicating that FRB 20201124A does not show regular burst intervals in the current dataset, consistent with the findings of \citet{2025arXiv250514219G}.

\subsection{Multi-band temporal offsets}

\begin{figure*}
\centering
\includegraphics[width=0.48\linewidth]{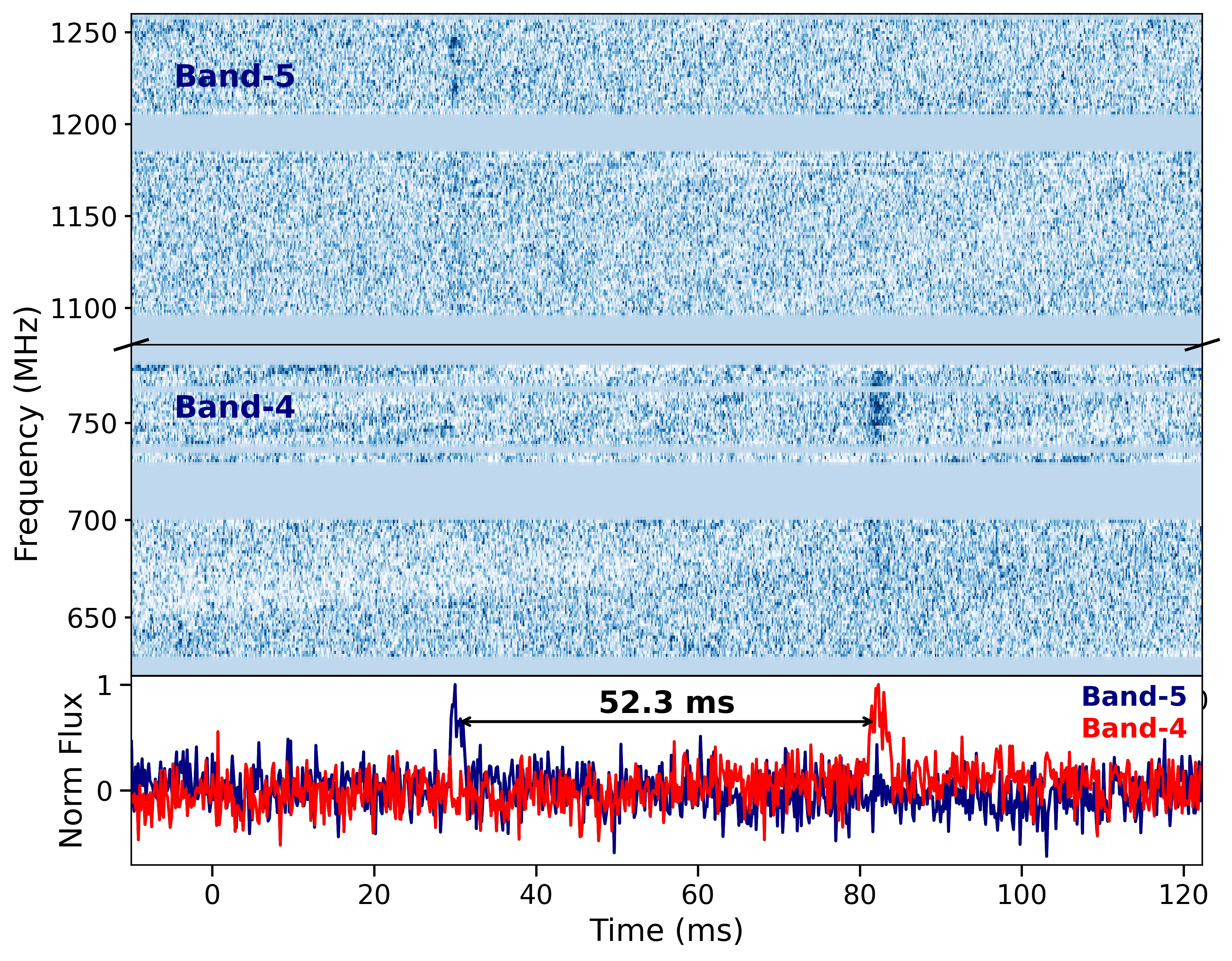}
\includegraphics[width=0.48\linewidth]{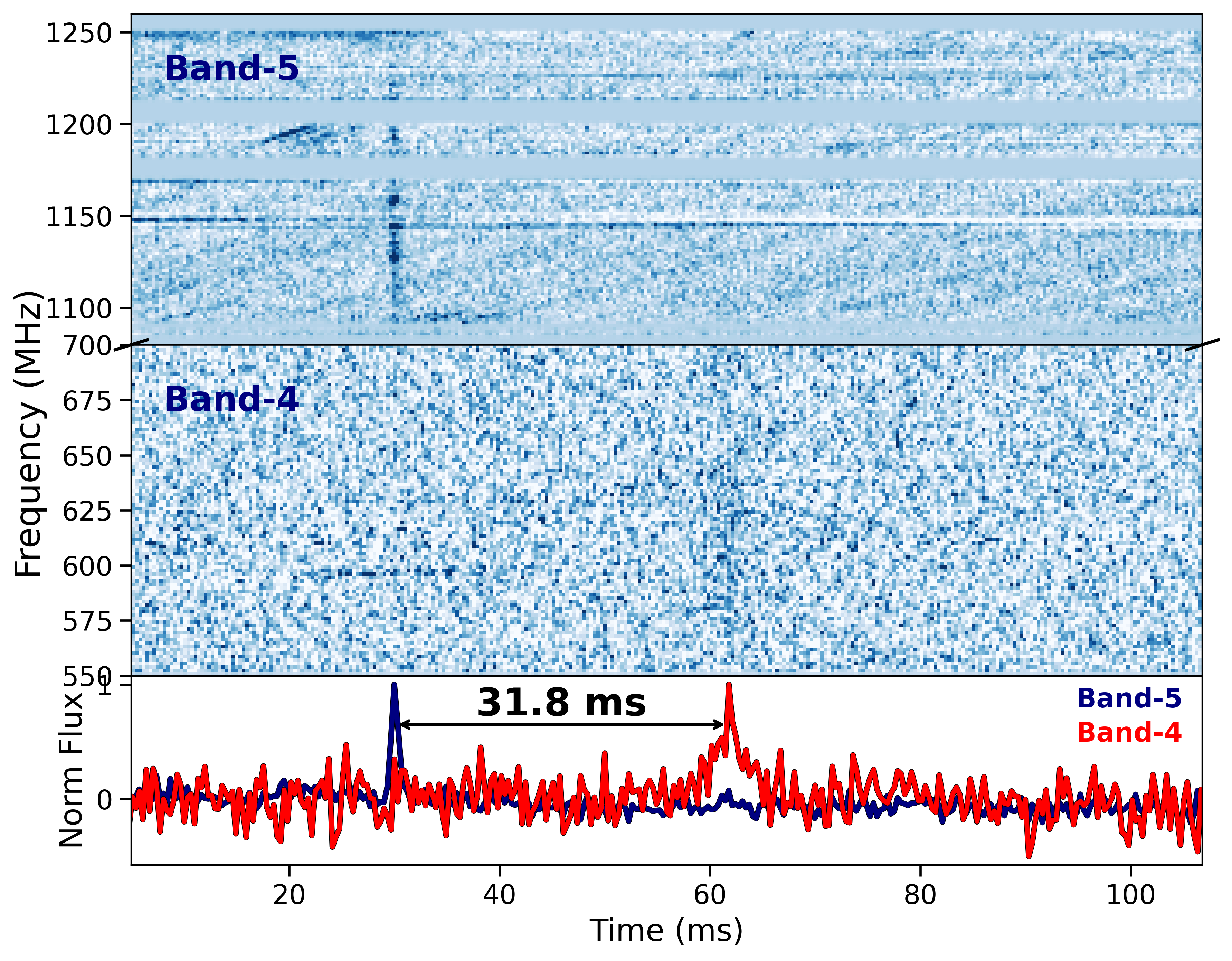}
\caption{
Simultaneous GMRT Band-4 (550--950 MHz) and Band-5 (1060--1460 MHz) detections of FRB~20201124A.
The upper and middle panels show the Band-5 and Band-4 dynamic spectra, respectively, while the bottom panel shows the frequency-averaged burst profiles.
On 9 May 2021 (left), the Band-5 burst precedes the Band-4 burst by 52.3\,ms.
On 18 May 2021 (right), the corresponding offset is 31.8\,ms.
These delays were measured after applying barycentric and inter-band dispersion corrections.}
\label{fig:multiband_offsets}
\end{figure*}

Simultaneous observations of FRB~20201124A were carried out using GMRT Band-4 (550--950 MHz) and Band-5 (1060--1460 MHz), allowing a direct comparison of burst arrival times across a wide frequency range. After applying barycentric corrections and correcting for inter-band dispersive delay, we identified two pairs of bursts detected in close temporal proximity during the observing sessions on 9 May and 18 May 2021.
\\
For the 9 May observation, a burst detected in Band-5 was found to precede a burst detected in Band-4 by $52.3$\,ms. Similarly, during the 18 May observation, a Band-5 burst arrived $31.8$\,ms earlier than a corresponding burst detected in Band-4. The dynamic spectra and frequency-averaged burst profiles for both events are shown in Figure~\ref{fig:multiband_offsets}.
\\
The uncertainties associated with the dispersion correction, obtained by propagating the uncertainties in the measured DM, are approximately 0.08~ms and 1.01~ms for the 9 May and 18 May observations, respectively. For bursts for which reliable profile fitting was not possible, we estimate the individual TOA uncertainties as half the effective boxcar width, giving uncertainties of 0.66~ms for both Band-4 and Band-5 on 9 May, and 2.62~ms and 0.66~ms for Band-4 and Band-5, respectively, on 18 May. These uncertainties are propagated through the barycentric conversion performed with \texttt{tempo2}. The measured offsets 52.3~ms and 31.8~ms are substantially larger than the estimated timing and dispersion-correction uncertainties. The observed offsets suggest that the emission detected in Band-4 and Band-5 was not strictly simultaneous, despite the observations being conducted simultaneously in sub-array mode.
\\
Interestingly, both measured delays are comparable to the short-timescale component of the waiting-time distribution shown in Figure~\ref{fig:4}, which peaks at $\sim$ 60 ms. This similarity suggests that the Band-4 and Band-5 burst pairs may be related to the same underlying short-timescale clustering behaviour observed in the source. In this interpretation, the events could represent closely spaced bursts rather than completely independent bursts.
\\
An alternative possibility is that the Band-4 and Band-5 bursts correspond to different frequency-dependent components of a single underlying burst. In the 9 May event, emission is detected up to $\sim$ 800 MHz in Band-4 and reappears near 1100 MHz in Band-5. However, the absence of frequency coverage between 800 and 1060 MHz prevents us from establishing whether the two detections are connected by continuous spectral evolution, potentially resembling the sad-trombone effect, or instead represent distinct bursts separated by a short time interval.
\\
Contemporaneous FAST observations provide additional context. The GMRT Band-5 bursts were detected simultaneously by FAST, consistent with the overlapping frequency coverage between GMRT Band-5 (1060$-$1460 MHz) and the FAST L-band (1050 MHz$-$1450 MHz) receiver. However, no strictly simultaneous counterparts were identified for the Band-4 bursts. Instead, the nearest GMRT Band-4--FAST burst pairs were separated by several seconds to tens of seconds. This behaviour suggests that the source activity is highly frequency dependent, with burst emission occupying different spectral regions at different times.
\\
Taken together, these results indicate that FRB~20201124A exhibits complex multi-frequency activity on timescales ranging from tens of milliseconds to hundreds of seconds. While the observed Band-4--Band-5 delays may reflect frequency-dependent burst structure, they are also consistent with the characteristic short-timescale clustering seen in the waiting-time distribution. The available data do not allow us to distinguish uniquely between these scenarios.

\subsection{Drift Rate}

\begin{figure}
    \centering
    \includegraphics[width=\linewidth]{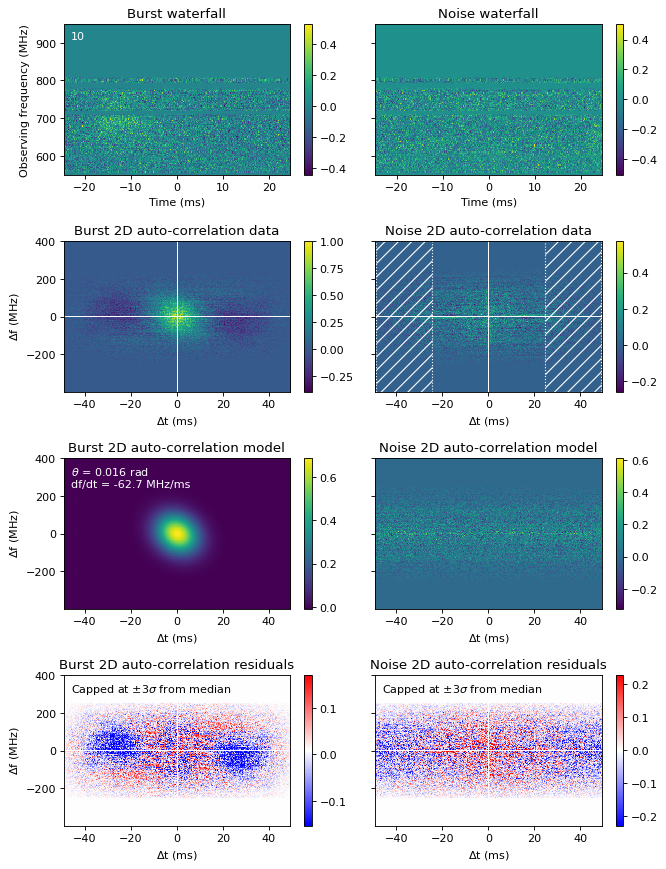}
    \caption{Auto-correlation drift rate analysis of a burst. 
        Top row: Waterfalls of the burst (left) and a noise region (right).
        Second row: 2D auto-correlation functions (ACFs) of both the burst and noise waterfalls, with the central peak at zero lag removed. The burst auto-correlation shows a clear diagonal elongation. Hatching indicates masked regions not used in modeling. 
        Third row: 2D Gaussian fit to the burst ACF (left), fixed at zero lag, yields a drift angle $\theta = 0.016$ rad and corresponding drift rate of $-62.7$ MHz ms$^{-1}$. The right panel shows the modeled 2D ACF of the noise, generated from per-$\Delta f$ channel statistics from the non-hatched region above. 
        Bottom row: Residuals between the data and models for burst (left) and noise (right) ACFs, clipped at $\pm 3\sigma$ from the median. Structured residuals in the burst indicate real substructure not captured by the simple Gaussian model, while noise residuals are consistent with uncorrelated fluctuations.}
    \label{fig 13}
\end{figure}

\begin{figure}
    \centering
    \includegraphics[width=\linewidth]{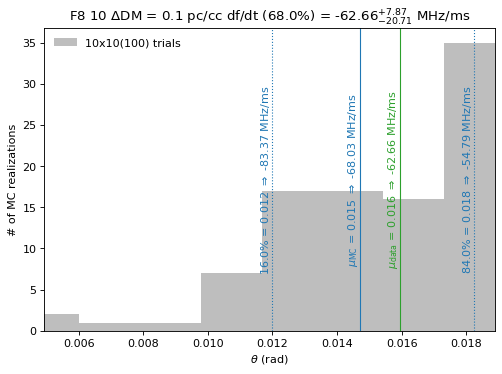}
    \caption{Monte Carlo distribution of the angle parameter ($\theta$) used to estimate the drift rate. The histogram shows $\theta$ values obtained from 100 resampling trials, with the corresponding drift rates overplotted. The central blue line marks the median $\theta$ value ($\mu_{\mathrm{MC}} = 0.015$~rad), corresponding to a drift rate of $-68.03$~MHz~ms$^{-1}$. The dashed blue lines represent the 68\% confidence interval ($\theta \in [0.012, 0.018]$), translating to drift rates from $-83.37$ to $-54.79$~MHz~ms$^{-1}$. The green line shows the mean drift rate over all realizations, $-62.66^{+7.87}_{-20.71}$~MHz~ms$^{-1}$. The distribution is asymmetric but excludes zero, indicating a statistically significant negative drift in this burst.
    }
    \label{fig 14}
\end{figure}

We estimated the drift rate of all bursts using \href{https://github.com/zpleunis/dfdt}{dfdt }\footnote{\url{https://github.com/zpleunis/dfdt}}\citep{Pleunis_2021}. This approach models the drift as the tilt of an elliptical shape in the two-dimensional autocorrelation of the dynamic spectrum. The linear drift rate \( \frac{d\nu}{dt} \) is related to the tilt angle \( \theta \) by

\[
\frac{d\nu}{dt} = \tan^{-1}(-\theta) \quad \text{MHz ms}^{-1}.
\]
The analysis begins by dedispersing the burst at a structure-optimising DM and extracting the waterfall, centred on the burst emission. Two-dimensional autocorrelations of both the burst and an off-burst (noise) region are calculated. The zero-lag peak, which can bias the measurement, is removed from these autocorrelations.
\\
A two-dimensional Gaussian function is fitted to the autocorrelation of the burst to determine the tilt angle $\theta$, which corresponds to the drift rate. To account for the effects of noise and uncertainty in DM, a Monte Carlo resampling approach is used. Multiple realisations of the noise autocorrelation model are generated and added to the burst autocorrelation before refitting the Gaussian to obtain a distribution of tilt angles.
\\
This process is repeated over a range of DMs sampled from the DM uncertainty distribution, allowing for a statistically robust estimate of the drift rate and its confidence intervals. The drift rate measurement is considered constrained if 99.73\% (3$\sigma$) of the sampled tilt angles share the same sign; otherwise, the drift rate is labelled unconstrained.
\\
Drift rates were calculated for all bursts, as summarised in the table given in the appendix section. Due to computational costs, the number of DM trials and Monte-Carlo trials was reduced to 10 for all the bursts. The results varied: some bursts showed positive drift rates, others had very large or near-zero values, indicating no clear drifting. However, for bursts affected by strong RFI, low brightness, or significant scattering, the drift rate estimates may not be reliable. Despite this, several measurements appear robust, while some remained unconstrained due to these limitations.

\section{Comparison with other sources}

\subsection{Cumulative Burst Rate}
We fitted the cumulative burst fluence distribution using a broken power-law, finding best-fit differential power-law indices of $\alpha_1 = -1.76 \pm 0.04$ and $\alpha_2 = -3.47 \pm 0.24$ using bursts above a fluence completeness limit of 3.05 Jy ms. These correspond to cumulative slopes of $-0.76 \pm 0.04$ and $-2.47 \pm 0.24$ below and above the break fluence, respectively. This indicates a relatively shallow slope at low fluences and a much steeper slope at higher fluences.
\\
For comparison, the CHIME/FRB Collaboration reported a differential fluence distribution with a power-law index of $\alpha = -4.6 \pm 1.3 \pm 0.6$ and a break at 16.6 Jy ms \citep{Lanman_2022}.
The other uGMRT data (based on less sensitive incoherent array) showed that the cumulative burst rate above a fluence completeness limit of 10 Jy ms follows a power-law with a cumulative index $\gamma = -1.2 \pm 0.2$ \citep{2022MNRAS.509.2209M}. Similarly, FAST observations reveal that the cumulative burst energy distribution is best modelled by broken power-law, with indices of $-0.22 \pm 0.01$ and $-3.27 \pm 0.23$ at low and high energies, respectively \citep{Zhang_2022}.
\\
Other repeating FRBs display comparable behaviour. FRB 20180916B exhibits a cumulative fluence distribution with a power-law index $\alpha = -2.3 \pm 0.3$ above 5.2 Jy ms \citep{2020Natur.582..351C}.  
\\
The steep high-fluence slope observed in FRB 20201124A is reminiscent of giant pulse energy distributions from the Crab pulsar. \citet{2019MNRAS.490L..12B} analysed over 1100 Crab giant pulses and found a power-law index near $-2.81 \pm 0.05$. MeerKAT observations of another young pulsar, PSR J0540$-$6919, yielded hundreds of GPs with a flux distribution that can be fitted by a power law with a very steep slope of $-2.75 \pm 0.02$ \citep{2021MNRAS.505.4468G}. 
\\
The similarity in power-law indices between FRB 20201124A and giant pulse distributions from young pulsars like the Crab and PSR J0540$-$6919 suggests that these systems may share common underlying mechanisms of coherent radio emission.

\subsection{Waiting time}


The waiting time distribution of FRB 20201124A displays a bimodal structure, with a short-timescale peak near 60 milliseconds, and a broader component around 97 seconds. Short waiting times of this scale are consistent with findings in other repeating FRBs. For example, FRB 20220912A shows a peak at $\sim$51 milliseconds \citep{Zhang_2023}, and an independent analysis of FRB 20201124A also reports a peak at 51.22 milliseconds \citep{Zhang_2022}. In FRB 20180916B, a characteristic short-timescale of 56.2 milliseconds was identified between closely spaced bursts during active windows \citep{2023MNRAS.524..569W}.
\\
These consistent millisecond-scale separations suggest that short waiting times are a common feature of active repeating FRBs. They likely reflect the temporal clustering of sub-bursts or physically distinct emission episodes occurring on rapid timescales. While the exact origin remains uncertain, such clustering may result from rapid variability in the emission region or modulation of coherent processes in the magnetosphere. The recurrence of this $\sim$50 ms scale across different sources points toward an intrinsic timescale in the emission mechanism or central engine dynamics common to repeaters, also pointed out by \citet{Zhang_2023}.
\\
While the short-timescale peak is broadly consistent across studies, the long-timescale component appears to vary between activity episodes. Our long-timescale peak of 97~s is consistent with that reported by \citet{Xu2022} from observations obtained during the May 2021 activity episode. Notably, despite the substantially different observing frequencies of FAST and uGMRT, both studies exhibit similar peak separations ($\sim$100s) and peak ratio ($\sim$ 6$-$7 times) of the two waiting-time components. In contrast, \citet{Zhang_2022}, using observations from a later activity episode in September 2021, reported a long-timescale peak near $\sim$10 s together with a different relative peak ratio of the two components ($\sim$1.6 times). This suggests that the observed differences are more likely related to changes in the source activity state than to observing frequency alone. Although sensitivity and frequency coverage can influence the detection of weaker bursts and closely spaced burst components, the agreement between the May 2021 FAST and uGMRT results, despite their substantially different frequency coverage, argues against frequency-dependent selection effects being the primary cause of the discrepancy. The relative number of bursts occurring at different separation timescales, particularly an excess of bursts separated by longer intervals, may indicate changes in the underlying emission state. Such characteristic separation timescales could potentially be associated with the build-up of magnetospheric twisting in a magnetar or the accumulation of stress in the neutron-star crust, eventually triggering burst emission.


\section{Discussion}
Our uGMRT observations provide a uniform, high-cadence view of FRB 20201124A during its active May 2021 episode. While several burst properties discussed in this work have been reported previously using observations from FAST, CHIME, GMRT, and other facilities, the simultaneous high-cadence multi-frequency coverage provided by the uGMRT allows a direct investigation of the frequency dependence of burst activity within a single observing framework. This reduces cross-instrument selection effects and provides a direct comparison of the source behaviour across a broad frequency range.
\\
One of the most notable results of this study arises from the simultaneous multi-frequency observations obtained with the uGMRT sub-array configuration. After applying barycentric and inter-band dispersion corrections, we identified pairs of bursts detected in Bands 4 and 5 that are separated by only 31.8 ms and 52.3 ms. These offsets are substantially larger than the uncertainties associated with the dispersion correction and are therefore unlikely to be instrumental in origin. Interestingly, both offsets are comparable to the short-timescale component of the waiting-time distribution, which peaks at $\sim$ 60 ms. This similarity suggests that the Band-4 and Band-5 burst pairs may be related to the same underlying short-timescale clustering behaviour observed in the source. The observed delays can be interpreted either as closely spaced bursts occurring within the same burst cluster or as frequency-dependent components of a single underlying event. In the 9 May observation (Figure~~\ref{fig:multiband_offsets}), emission is detected up to $\sim$ 800 MHz in Band-4 and reappears near 1100 MHz in Band-5, although the absence of frequency coverage between 800 and 1060 MHz prevents us from determining whether the emission evolves continuously across the intervening frequencies. Regardless of the interpretation, these observations demonstrate that burst activity across different frequency bands is not always strictly simultaneous and highlight the complex frequency-dependent behaviour of FRB~20201124A.
\\
Comparison with contemporaneous FAST L-band \citep{Xu2022} and uGMRT Band-4 observations reveals additional burst activity occurring within a few to several tens of seconds, although no strict one-to-one correspondence is observed between individual bursts. Taken together, these observations suggest that burst activity across different frequency bands is not always strictly simultaneous and are consistent with frequency-dependent burst occurrence and patchy spectral occupancy. In this picture, individual bursts do not necessarily detected across the full observable radio band at a given time. The simultaneous multi-frequency coverage provided by the uGMRT therefore offers an independent perspective on the broadband activity of FRB 20201124A and complements previous multi-frequency studies of this source.
\\
A second notable result is the frequency-dependent evolution of the activity window. We find that Band-5 activity ceased on 24 May 2021, whereas Band-4 remained active until 28 May 2021. This indicates that the higher-frequency emission shut off earlier while lower-frequency activity persisted for several additional days. Similar chromatic activity evolution has been reported previously by \citet{Kirsten2024} and \citet{2025A&A...696A.194B}. Our observations provide an independent confirmation of this behaviour within a uniform observing setup and further support the picture that the activity window of FRB 20201124A evolves with frequency.
\\
We also detected closely spaced burst pairs, with separations as short as $\sim$ 17 ms. Such short intervals imply that the emission arises from compact regions where magnetospheric processes operate on millisecond timescales. While we cannot confirm strict periodicity, the presence of sub-second repetition is consistent with other repeaters that show quasi-periodic or clustered behaviour \citep{Pastor, Faber}. Together with the bimodal waiting-time and energy distributions in our sample, these results point to a complex burst-generation process involving multiple characteristic timescales. The absence of a significant correlation between burst energy and waiting time suggests that the observed energy and temporal distributions are not directly linked. 
The consistency of the $\sim$100 s waiting-time component across widely separated observing frequencies, together with its variation between activity epochs, suggests that the long-timescale burst clustering is primarily governed by changes in the intrinsic activity state of the source rather than frequency-dependent selection effects. These characteristic timescales may trace the build-up and release of magnetospheric or crustal stress, providing a potential link between the observed burst statistics and the underlying magnetar activity cycle.
\\
These findings naturally connect to the complex magneto-ionic environment of the source. FRB 20201124A shows strong and variable polarisation \citep{Jiang2024, Hilmarsson_2021} and large RM variations \citep{Xu2022}, pointing to changes in the magnetic field and plasma density around the source. Polarisation studies also suggest the presence of multiple active emission regions within the magnetosphere. Our detection of dual timescales and dual energy modes may be the temporal signature of these different regions switching on and off.
\\
Finally, the cumulative burst fluence distribution is well described by a broken power law, similar to what has been found in other repeaters. This suggests the view that repeating FRBs share a common underlying energy release process, even if their activity cycles and frequency dependence differ.
\\
Overall, our results provide a simultaneous multi-frequency view of FRB 20201124A during its May 2021 activity episode. The observed Band-4--Band-5 delays of 31.8 ms and 52.3 ms, together with the frequency-dependent evolution of the activity window, demonstrate that burst activity is not always observed simultaneously across a broad radio band. Combined with contemporaneous FAST observations, these results suggest that the source exhibits complex frequency-dependent burst behaviour and patchy spectral occupancy. More generally, our findings highlight the importance of uniform, high-cadence, multi-frequency observations for understanding the temporal and spectral properties of repeating FRBs and for distinguishing between intrinsic burst evolution and frequency-dependent emission processes.

\section{Summary}

We conducted multi-epoch uGMRT observations of FRB 20201124A between 8 May and 15 June 2021, covering 300–1460 MHz. A total of 146 bursts were detected, predominantly in Band 4 (550--950 MHz), providing a high-cadence multi-frequency view of the source during its active May 2021 episode. A notable result of this study is the identification of closely spaced burst events in the simultaneously observed Band-4 and Band-5 data, with temporal offsets of 31.8 ms and 52.3 ms after applying barycentric and inter-band dispersion corrections. These delays are comparable to the short-timescale component of the waiting-time distribution, which peaks at 60 ms, suggesting a possible connection between the observed multi-band burst pairs and the short-timescale clustering behaviour of the source. The Band-5 bursts were also detected contemporaneously by FAST, whereas no strictly simultaneous FAST counterparts were identified for the corresponding Band-4 bursts. Together, these observations suggest that burst activity across different frequency bands is not always strictly simultaneous and are consistent with frequency-dependent burst occurrence and patchy spectral occupancy. We also note the epoch-dependent variation of the long-timescale waiting-time component likely reflecting intrinsic changes in the source activity state, potentially tracing the build-up and release of magnetospheric or crustal stress. We see frequency-dependent activity cessation with Band-5 activity ceasing few days before Band-4 activity. This independently confirms the chromatic activity window previously reported for FRB 20201124A using observations from multiple facilities. Several sub-second burst pairs were observed with separations as short as 17 ms. The burst population exhibits bimodal waiting-time and energy distributions, indicating at least two characteristic emission timescales and energy modes. Dispersion measures span the range $\sim$ 410--420 pc cm$^{-3}$, with most bursts clustered near 412--415 pc cm$^{-3}$. Intrinsic burst widths are typically below 10 ms, scattering timescales are generally less than 5 ms, and the cumulative fluence distribution is well described by a broken power law with a break near $\sim$17 Jy ms. Overall, our results provide a uniform multi-frequency view of FRB 20201124A during its May 2021 active episode. The observed frequency-dependent burst behaviour, together with the non-coincident activity seen across different frequency bands, highlights the importance of simultaneous wideband observations for understanding the emission properties of repeating FRBs.

\section{ACKNOWLEDGEMENTS}

We acknowledge the support of the Department of Atomic Energy, Government of India, under project title ``Next Generation Instrumentation for Radio Astronomy" No. RTI4017. We thank the uGMRT operators for their coordinated effort in conducting the DDT observations. The GMRT is run by the National Centre for Radio Astrophysics of the Tata Institute of Fundamental Research, India. We also thank Heng Xu, Suryarao Bethapudi, and Arpan Pal for helpful discussions. We acknowledge Nissim Kanekar and Aditya Chowdhury for their contributions to the original uGMRT DDT proposals, which enabled the observations of FRB 20201124A. We thank the anonymous referee for their comments and suggestions, which helped to improve the paper.

\section{SOFTWARES}
\facilities{Giant Meterwave Radio Telescope, Khodad, Pune}

\software{ gptool \citep{gptool}, SIGPROC \citep{sigproc}, PRESTO \citep{2001PhDT.......123R}, your \citep{Aggarwal2020}, scipy \citep{2020NaMet..17..261V}, scikit-learn \citep{JMLR:v12:pedregosa11a}, matplotlib \citep{2007CSE.....9...90H}}

\bibliography{references}{}
\bibliographystyle{aasjournalv7}

 \appendix

 \section{Burst detection plots for FRB 20201124A}
\begin{figure}[htbp]
    \centering
    \includegraphics[height=0.8\textheight, width =\textwidth]{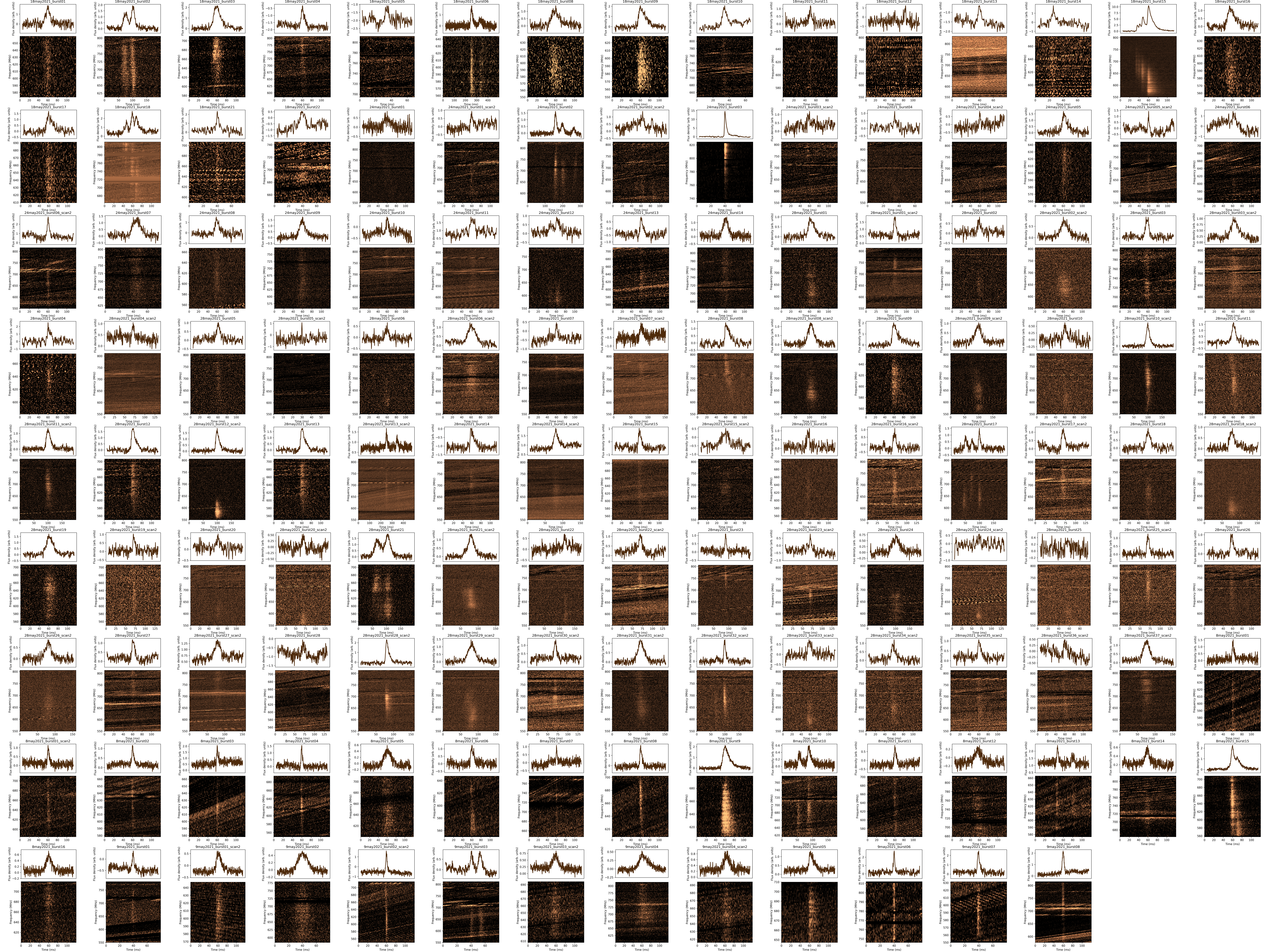}
    \caption{Burst detection plots from FRB 20201124A detected in the uGMRT Band-4 observations. Each panel corresponds to an individual burst and displays the burst profile and dynamic spectrum.
}
    \label{fig:appendix_burst}
\end{figure}

\section{BURST PROPERTIES }

\scriptsize
\setlength{\tabcolsep}{0.5pt}

{\tiny
\renewcommand{\arraystretch}{1}
{\fontsize{2}{3}\selectfont
\begin{longtable}{lccccccccc}

\hline

MJD & S/N & DM & $S_{\rm peak}$ & $W_{\rm int}$ & {$\tau_{\rm scat}$}  & Fluence & $\nu_{\rm peak}$ & BW & Drift Rate
\\
\hline
59342.2934007694(20) & 12.16 & 411.69(22) & 0.87 & 2.49(46) & 3.57(51)  & 3.77(53) &  &  & $0.10$ \\
\hline
59342.2970455001(25) & 23.36 & 415(18) & 1.55 & 5.79(50) & 4.51(49)  & 11.41(91) &  &  & $-28.16^{+1.6}_{-3.7}$ \\
\hline
59342.2996973389(11) & 8.56 & 414(7) & 0.63 & 1.78(24) & 2.08(26)  & 1.72(26) & 684.0(24) & 127.2(73) & $0.35^{+0.3}_{-0.3}$ \\
\hline
59342.307094207(8) & 15.82 & 411.64(9) & 1.16 & 1.99(17) & 1.69(16)  & 3.02(27) & 745.4(25) & 69.5(66) & $-22.46^{+1.2}_{-2.4}$ \\
\hline
59342.310627178(63) & 30.62 & 416.30(26) & 1.38 & 24.6(10) & 3.0(11)  & 34.3(18) & 632.0(60) & 142(18) & $-60.40$ \\
\hline
59342.3128994328(23) & 16.08 & 411.69(16) & 1.08 & 3.80(52) & 5.87(63)  & 7.58(80) &  &  & $0.32$ \\
\hline
59342.3152898374(28) & 11.53 & 412.61(19) & 0.84 & 2.62(55) & 2.21(55)  & 2.87(52) &  &  & $-11.01^{+1.5}_{-0.8}$ \\
\hline
59342.3155331427(9) & 15.78 & 412.32(30) & 1.13 & 2.62(19) & 2.94(20)  & 4.46(36) & 758.7(41) & 133(13) & $0.11^{+0.1}_{-0.0}$ \\
\hline
59342.3155360336(9) & 41.19 & 413.18(32) & 2.25 & 8.02(20) & 13.14(25)  & 34.63(100) & 708.0(31) & 132(19) & $-14.45^{+0.2}_{-0.0}$ \\
\hline
59342.3189691316(6) & 22.03 & 413.60(26) & 1.62 & 1.81(18) & 3.19(43)  & 5.93(68) &  &  & $-65.90^{+6.9}_{-14.5}$ \\
\hline
59342.3201042803(28) & 15.73 & 414.64(12) & 1.11 & 3.77(50) & 2.42(48)  & 5.00(63) &  &  & $-9.54^{+2.2}_{-5.5}$ \\
\hline
59342.3269777749(132) & 25.27 & 417.64(14) & 1.29 & 17.3(16) & 6.0(17)  & 23.6(23) & 663.4(40) & 44.6(99) & $-23.90$ \\
\hline
59342.3274120113(44) & 12.86 & 414.41(13) & 0.94 & 2.51(74) & 1.37(71)  & 2.69(72) &  &  & $0.08^{+0.0}_{-0.0}$ \\
\hline
59342.3275322272(23) & 20.47 & 411.67(34) & 1.48 & 3.60(39) & 0.9(26)  & 5.5(12) & 680.6(64) &  & $-19.34$ \\
\hline
59342.3281808499(60) & 50.87 & 410.64(5) & 2.49 & 19.39(60) & 6.35(65)  & 50.9(18) & 639(11) & 175.5(56) & $-12.60$ \\
\hline
59342.3322539638(2) & 20.63 & 412.10(0) &  & 12.9(13) & 7.1(13)  &  & 615.48(98) & 110.1(26) &  \\
\hline
59342.3492221554(5) & 14.23 & 413.47(24) & 1.02 & 3.77(50) & 1.45(50)  & 4.11(59) &  &  & $-27.29^{+4.6}_{-11.5}$ \\
\hline
59343.2220587643(21) & 14.84 & 414.80(12) & 1.08 & 2.64(36) & 1.64(35)  & 3.35(45) &  &  & $-2.01$ \\
\hline
59343.2315640607(22) & 32.49 & 417.07(17) & 2.11 & 7.61(32) & 3.21(32)  & 17.46(86) &  &  & $-482.46$ \\
\hline
59343.2436146812(19) & 15.87 & 415.10(64) & 1.14 & 2.81(39) & 2.48(38)  & 4.27(51) &  &  & $-49.25$ \\
59343.2436148752(9) &  &  & 1.13 & 4.198(60) &   &  &  &  &  \\
\hline
59343.2481925062(21) & 31.20 & 418.88(63) & 1.84 & 9.31(40) & 7.85(41)  & 22.4(10) &  &  & $-72.94^{+5.4}_{-11.4}$ \\
\hline
59343.2524692863(17) & 22.37 & 411.95(26) & 1.52 & 5.37(32) & 3.84(31)  & 10.04(66) & 575.3(48) &  & $-2.33^{+0.1}_{-0.1}$ \\
\hline
59343.2531710926(9) & 18.35 & 415.15(3) & 1.35 & 2.18(11) & 0.224(84)  & 2.95(22) &  &  & $-240.79^{+132.7}_{-659.3}$ \\
\hline
59343.2548053149(8) & 16.06 & 414.42(27) & 1.17 & 2.11(18) & 2.35(18)  & 3.69(31) &  &  & $-4.00$ \\
\hline
59343.2551530509(48) & 37.88 & 416.6(16) & 2.01 & 15.29(67) & 6.53(69)  & 33.4(16) & 800.1(27) & 89.8(73) & $0.26^{+0.3}_{-0.3}$ \\
\hline
59343.2791084002(41) & 21.30 & 413.73(16) & 1.55 & 2.91(17) & 1.94(16)  & 5.41(36) & 697.9(50) & 143(13) & $-6.18^{+0.8}_{-0.6}$ \\
\hline
59343.2914789871(15) & 9.47 & 412.80(7) & 0.67 & 2.54(35) & 3.52(39)  & 2.93(40) & 674.6(67) & 104(19) & $0.50^{+0.1}_{-0.1}$ \\
\hline
59343.2938063368(82) & 23.47 & 414.59(10) & 1.33 & 13.14(99) & 4.3(10)  & 18.4(15) & 724.6(32) & 91(11) & $-3.28^{+0.1}_{-0.0}$ \\
\hline
59343.3047088206(7) & 20.76 & 411.66(9) & 1.51 & 2.06(16) & 2.60(17)  & 5.02(35) & 630.7(12) & 112.4(28) & $-11.79^{+0.2}_{-0.4}$ \\
\hline
59352.2718159328(110) & 17.39 & 417.58(41) & 1.11 & 10.2(13) & 3.1(13)  & 11.9(16) &  &  & $-13.36$ \\
\hline
59352.2726995011(19) & 22.92 & 414.72(18) & 1.37 & 8.17(43) & 10.62(64)  & 18.4(11) &  &  & $-39.42^{+0.6}_{-4.1}$ \\
\hline
59352.2761271669(22) & 32.65 & 415.73(10) & 1.81 & 11.93(40) & 8.47(41)  & 26.5(11) & 784.19(32) & 140.00(65) & $0.00$ \\
\hline
59352.2766297656(74) & 18.76 & 414.76(6) & 1.20 & 8.91(74) & 1.29(63)  & 10.8(11) & 663.6(63) & 179(21) & $-0.41$ \\
\hline
59352.27708038700 & 6.91 & 409.75(10) &  &  &   &  &  &  & $1.95$ \\
\hline
59352.2775588204(199) & 26.32 & 413.68(13) & 1.14 & 24.9(36) & 22.2(37)  & 38.2(44) &  &  & $-31.62^{+2.1}_{-7.5}$ \\
59352.2775597686(39) & 26.32 &  & 1.40 & 12.27(85) &   &  &  &  &  \\
\hline
59352.2795756577(34) & 21.83 & 413.72(15) & 1.43 & 7.09(68) & 6.75(71)  & 14.0(12) & 652.5(76) & 142(23) & $-10.67$ \\
\hline
59352.2810116118(23) & 31.94 & 412.71(12) & 1.80 & 5.60(57) & 15.06(87)  & 28.9(18) & 620.5(42) & 151(12) & $29.72$ \\
\hline
59352.2823821356(189) & 14.73 & 415.53(4) & 1.02 & 5.5(18) & 1.3(21)  & 5.8(19) &  &  & $-5.92$ \\
\hline
59352.283235482(69) & 9.41 & 415.23(38) & 0.69 & 4.5(12) & 2.49(65)  & 3.57(83) &  &  & $-6.64$ \\
\hline
59352.28515949160 & 8.81 & 414.83(45) &  &  &  &   &  &  & $0.24$ \\
\hline
59352.2853334957(82) & 10.75 & 414.75(9) & 0.72 & 6.7(10) & 1.8(11)  & 5.06(87) & 638.4(36) & 176(12) & $0.28$ \\
\hline
59352.2934026934(95) & 12.90 & 414.78(17) &  &   &  &  &  &  & $-4.89$ \\
\hline
59352.2982900938(8) & 56.36 & 410.82(12) & 2.33 & 17.99(91) & 28.5(13)  & 78.5(32) & 603.8(54) & 210.4(42) & $-5.70$ \\
\hline
59352.3012724544(211) & 22.51 & 413.94(2) & 1.55 & 5.08(53) & 6.03(57)  & 12.2(10) & 645(14) & 127(52) & $-0.71$ \\
\hline
59352.3013166214(177) & 20.29 & 410.11(12) & 1.48 & 3.68(45) & 4.47(48)  & 8.55(81) & 698(11) & 75(29) & $-8.00$ \\
\hline
59352.3055189081(35) & 34.63 & 412.47(8) & 2.47 & 3.31(72) & 10.2(15)  & 26.6(37) & 757.34(96) & 138.4(24) &  \\
59352.3055193258(19)   &  & & 2.53 & 3.06(49) &  &   &  &  &  \\
59352.305519489(26)   &  & & 1.93 & 10.923(51) &  &    &  &  &  \\
\hline
59352.3075182278(19) & 11.44 & 411.83(20) & 0.86 & 2.86(48) & 2.75(48)  & 3.42(51) & 694.8(10) &  & $-42.77$ \\
\hline
59352.3081496717(45) & 12.83 & 413.66(10) & 0.79 & 10.32(13) & 1.231(15)  & 8.25(65) &  &  & $8.33$ \\
\hline
59358.2420434049 & 11.75 & 412.46(82) &  &  &  &   &  &  & $0.13$ \\
\hline
59358.244447671(24) & 23.86 & 416.70(7) & 1.14 & 22.2(24) & 9.3(25)  & 27.6(30) & 720.9(80) & 56(20) & $71.18^{+7.3}_{-10.3}$ \\
\hline
59358.2444508922(2) & 66.57 & 410.72(11) & 4.93 & 1.417(45) & 1.355(60)  & 9.67(30) & 821.1(32) & 80.2(85) & $-13.80$ \\
\hline
59358.2449155718(32) & 10.04 & 410.75(14) &  &  &   &  &  &  & $-38.16$ \\
\hline
59358.2550475529(28) & 23.59 & 412.68(9) & 1.66 & 6.92(69) & 1.14(76)  & 11.7(13) & 622.0(69) & 94(19) & $-34.40$ \\
\hline
59358.26176732450 & 11.19 & 411.76(22) &  &  &   &  &  &  & $-13.95$ \\
\hline
59358.2621983584(57) & 25.53 & 413.65(9) & 1.47 & 12.42(73) & 4.46(76)  & 19.4(13) &  &  & $-177.46$ \\
\hline
59358.26465137050 & 18.27 & 416.75(22) &  &  &   &  &  &  & $-28.34$ \\
\hline
59358.2685162952(28) & 30.06 & 415.47(13) & 1.72 & 10.08(54) & 8.97(57)  & 23.2(12) & 665.92(46) &  & $0.00$ \\
\hline
59358.27196901210 & 10.56 & 413.68(8) &  &  &   &  &  &  & $-541.46$ \\
\hline
59358.27507193010 & 13.33 & 419.83(4) &  &  &  &  &  &  &  \\
\hline
59358.27707494190 & 18.57 & 415.81(19) &  &  &   &  &  &  & $-12.17$ \\
\hline
59358.27759291410 & 12.30 & 414.27(28) &  &  &  &   &  &  & $0.13$ \\
\hline
59358.3008676366(168) & 24.29 & 416.53(45) & 1.37 & 13.9(11) & 2.0(17)  & 19.2(17) & 752.8(16) & 50.6(38) & $133.57$ \\
\hline
59358.30360502360 & 8.38 & 413.30(10) &  &  &  &   &  &  &  \\
\hline
59358.30364204850 & 11.14 & 410.69(20) &  &  &   &  &  &  & $0.32$ \\
\hline
59358.30372522200 & 11.24 & 412.18(6) &  &  &  &   &  &  & $-0.12$ \\
\hline
59358.30978049650 & 6.96 & 415.44(36) &  &  &   &  &  &  & $-0.12$ \\
\hline
59358.31243151220 & 7.59 & 418.94(60) &  &  &   &  &  &  & $0.94$ \\
\hline
59358.312683702(169) & 16.65 & 413.88(11) & 1.03 & 10.0(11) & 1.8(17)  & 10.5(14) &  &  & $43.34$ \\
\hline
59362.1885058265(34) & 27.48 & 415.15(15) & 1.63 & 11.05(63) & 8.36(65)  & 22.6(13) & 665.4(51) & 50(13) & $-28.49$ \\
\hline
59362.1896390285(39) & 10.29 & 412.31(18) & 0.72 & 4.67(86) & 5.54(92)  & 5.24(82) &  &  & $-28.28$ \\
\hline
59362.1921487417(21) & 11.28 & 419.46(1) & 0.75 & 8.88(38) & 1.20(13)  & 6.76(66) &  &  & $-5.96$ \\
\hline
59362.1926074876(38) & 13.99 & 413.13(4) & 0.96 & 4.88(85) & 6.33(94)  & 7.7(10) & 779(14) & 155(50) & $-170.44$ \\
\hline
59362.1932796555(54) & 11.61 & 415.04(26) & 0.78 & 8.86(57) & 1.01(46)  & 6.95(74) &  &  & $-29.96$ \\
\hline
59362.1942600119(16) & 12.83 & 411.72(52) & 0.86 & 9.112(30) & 1.282(10)  & 7.87(61) & 658.9(54) & 66(14) & $-142.73$ \\
\hline
59362.1948202805(56) & 12.90 & 412.58(6) & 0.90 & 4.77(91) & 2.54(88)  & 4.86(90) &  &  & $0.08$ \\
\hline
59362.1954948093(15) & 13.78 & 412.35(25) & 0.96 & 4.59(34) & 5.91(37)  & 7.17(62) & 744(14) & 121(35) & $-14.80$ \\
\hline
59362.1983085733(18) & 32.84 & 411.31(12) & 2.02 & 5.34(44) & 11.31(59)  & 25.2(14) & 649.7(54) & 174(18) & $-18.63$ \\
\hline
59362.19934108470 & 12.07 & 412.46(10) &  &   &  &  &  &  & $0.03$ \\
\hline
59362.2013454594(28) & 18.86 & 413.67(8) & 1.29 & 7.18(47) & 3.97(45)  & 10.59(82) &  &  &  \\
\hline
59362.2035708785(13) & 27.74 & 412.30(5) & 1.84 & 4.81(31) & 8.10(37)  & 17.37(91) &  &  & $0.03$ \\
\hline
59362.2048295149(35) & 22.06 & 414.40(145) & 1.43 & 9.16(55) & 4.58(54)  & 14.6(10) & 753.88(59) & 44.6(15) & $145.29^{+45.7}_{-62.4}$ \\
\hline
59362.2053585992(14) & 14.56 & 414.39(12) & 0.97 & 2.58(36) & 7.12(51)  & 7.31(69) &  &  & $-85.12$ \\
\hline
59362.2070272273(29) & 15.57 & 414.48(48) & 1.05 & 5.04(59) & 4.91(61)  & 7.36(79) & 683.2(31) & 56.0(79) & $0.22$ \\
\hline
59362.2096851003(139) & 11.88 & 412.95(78) & 0.85 & 6.4(13) & 1.4(15)  & 5.5(12) & 719.0(66) & 41(16) & $0.00$ \\
\hline
59362.2100193241(68) & 14.78 & 415.44(4) & 1.08 & 4.7(12) & 3.1(11)  & 6.1(13) &  &  & $71.80$ \\
59362.2100196621(29)  &  & & 1.05 & 6.759(33) &  &  &    &  &  \\
59362.2100199661(5)  &  & & 1.09 & 4.830(20) &  &  &    &  &  \\
\hline
59362.2121951954(38) & 16.31 & 416.63(10) & 1.07 & 8.07(69) & 5.49(68)  & 10.47(98) &  &  &  \\
\hline
59362.2135638178(57) & 32.14 & 416.34(9) & 1.79 & 15.52(76) & 6.02(78)  & 29.7(16) & 761.3(54) & 84(14) & $-118.53$ \\
\hline
59362.2138715249(62) & 11.23 & 417.18(21) & 0.77 & 5.3(13) & 6.3(14)  & 6.3(12) & 694(16) & 138(52) & $0.26$ \\
\hline
59362.2147772334(66) & 29.35 & 417.71(12) & 1.41 & 18.3(11) & 17.5(20)  & 35.7(25) & 688.8(41) & 74(11) & $-5.45^{+1.3}_{-1.0}$ \\
59362.2147777394(55) & &  & 1.63 & 15.08(94) &  &  &  &  &  \\
\hline
59362.2161708355(46) & 12.12 & 412.78(7) & 0.85 & 5.57(90) & 4.65(78)  & 6.17(89) &  &  & $77.72$ \\
\hline
59362.2165086534(21) & 6.98 & 412.13(7) & 0.53 & 3.35(38) & 2.14(36)  & 2.09(36) & 737.3(78) & 139(27) & $-67.68^{+24.9}_{-79.7}$ \\
\hline
59362.21926939770 & 24.74 & 417.66(3) & 1.50 & 12.8(10) & 1.249(40)  & 19.3(17) & 730(20) &  & $145.29^{+45.7}_{-0.0}$ \\
\hline
59362.22244723220 & 9.33 & 411.12(49) &  &  &    &  &  &  & $-8.27^{+2.3}_{-8.8}$ \\
\hline
59362.22404810800 & 11.87 & 415.49(64) &  &  &  &  &  &  &  \\
\hline
59362.2264377002(44) & 16.25 & 415.14(10) & 1.12 & 5.35(65) & 2.39(64)  & 6.54(83) &  &  & $0.12$ \\
\hline
59362.22754473170 & 7.87 & 417.92(98) &  &  & 
&  &  &  & $-5.68$ \\
\hline
59362.2467795635(37) & 13.67 & 412.55(2) & 0.95 & 5.29(48) & 1.94(49)  & 5.35(61) &  &  & $0.15$ \\
\hline
59362.2469517594(56) & 22.38 & 415.95(18) & 1.52 & 8.29(81) & 1.23(83)  & 12.8(14) & 684.5(55) & 81(15) & $-14.36$ \\
\hline
59362.2476674375(46) & 20.44 & 417.75(24) & 1.29 & 8.68(69) & 4.01(68)  & 12.3(11) & 786.8(22) & 151.9(45) &  \\
\hline
59362.24775304890 & 7.64 & 411.77(11) &  &  &  &    &  &  & $0.19$ \\
\hline
59362.25014487020 & 9.00 & 414.72(4) &  &  &  &    &  &  &  \\
\hline
59362.2508497263(60) & 22.21 & 417.10(11) & 1.42 & 8.62(65) & 2.37(71)  & 12.7(11) & 672.9(33) & 120.9(81) &  \\
\hline
59362.25103010860 & 11.61 & 416.69(14) &  &  &    &  &  &  &  \\
\hline
59362.2513555656(48) & 29.45 & 420.22(80) & 2.10 & 6.22(60) & 0.94(65)  & 13.2(13) & 680.6(91) & 178(33) &  \\
\hline
59362.2525187689(32) & 24.57 & 414.87(36) & 1.69 & 7.62(33) & 1.727(61)  & 13.25(77) & 713.806 & 120.516 &  \\
\hline
59362.2540939024(16) & 40.94 & 415.21(5) & 2.95 & 5.66(84) & 1.80(84)  & 17.6(25) & 773.4(36) & 137(12) &  \\
\hline
59362.2546079067(20) & 26.15 & 415.1(15) & 1.77 & 3.7(11) & 7.7(11)  & 15.2(20) & 728.6(63) & 305(93) &  \\
\hline
59362.2546114324(9) & 50.46 & 413.33(65) & 3.45 & 8.13(60) & 1.58(66) & 28.5(22) & 715.9(36) & 212(20) &  \\
\hline
59362.2552966942(16) & 22.47 & 414.76(8) & 1.52 & 4.13(35) & 5.21(37)  & 10.14(71) & 669.5(73) & 130(22) &  \\
59362.2552973462(25)  &  & & 1.41 & 9.805(69) &  &   &  &  &  \\
\hline
59362.256485245(19) & 20.10 & 414.51(19) & 0.83 & 13.3(14) & 27.3(20)  & 25.2(20) & 642.0(33) & 146(10) &  \\
\hline
59362.2584494286(24) & 12.47 & 416.77(136) & 0.79 & 9.30(46) & 1.354(71)  & 7.40(69) &  &  &  \\
\hline
59362.2586227722(78) & 13.67 & 414.80(39) & 0.70 & 17.3(10) & 6.4(14)  & 12.9(12) &  &  & $-9.72$ \\
\hline
59362.2596954728(57) & 14.59 & 414.65(6) & 0.95 & 7.46(88) & 3.56(86)  & 7.86(99) &  &  & $-206.79$ \\
\hline
59362.2623677142(24) & 36.78 & 420.56(1) & 2.10 & 10.72(50) & 11.33(55)  & 32.7(14) & 711.0(95) & 206(48) &  \\
\hline
59362.26395159400 & 11.14 & 412.63(8) &  &  &    &  &  &  & $-0.03$ \\
\hline
59362.2671315887(40) & 12.71 & 417.70(3) & 0.86 & 5.88(83) & 5.99(87)  & 7.25(93) &  &  & $51.64$ \\
\hline
59362.26734801930 & 28.28 & 412.79(53) & 1.62 & 15.37(27) & 1.690(17)  & 25.00(98) &  &  & $0.18$ \\
\hline
59362.2679915506(2) & 13.01 & 414.41(12) & 0.91 & 5.00(16) & 1.082(30)  & 4.67(39) & 713.7(20) & 55.1(51) & $0.18$ \\
\hline
59362.2709263754(2) & 13.10 & 412.72(18) &  &    &  &  &  &  &  \\
\hline
59362.27151149(106) & 15.62 & 418.24(18) &  &    &  &  &  &  &  \\
\hline
59362.2715658906(37) & 10.41 & 415.69(79) & 0.74 & 5.94(61) & 3.31(59)  & 5.00(66) & 742.000 & 47.000 &  \\
\hline
59362.271636863(150) & 21.51 & 419.13(22) & 1.32 & 12.2(10) & 1.7(15)  & 16.3(16) & 772.5(51) & 56(13) &  \\
\hline
59362.271984599(34) & 20.22 & 417.75(52) & 1.43 & 4.35(51) & 1.27(50)  & 6.48(80) & 802.17(67) & 90.2(16) &  \\
\hline
59362.2725319007(6) & 45.51 & 410.63(8) & 2.95 & 4.76(14) & 7.16(16)  & 25.33(72) &  &  &  \\
\hline
59362.2731576653(66) & 34.04 & 418.84(12) & 1.43 & 32.16(76) & 2.96(84)  & 46.3(17) & 677.6(35) & 46.3(88) &  \\
\hline
59362.2738441513(32) & 13.25 & 415.63(8) & 0.90 & 4.88(63) & 4.15(62)  & 5.80(72) &  &  &  \\
\hline
59362.2742322747(6) & 27.37 & 416.34(25) & 1.68 & 12.277(32) & 1.272(30)  & 20.78(76) &  &  &  \\
\hline
59362.2745336322(30)  & 19.95 & 412.59(4) & 1.29 & 1.80(81) & 11.5(16)  & 15.0(22) & 745.2(83) & 183(43) &  \\
59362.2745340301(11)  &  &  & 1.40 & 4.47(24) &  &  &  &  &  \\
\hline
59362.2767750596(41) & 17.76 & 414.80(12) & 1.21 & 6.90(76) & 4.81(74)  & 10.2(11) & 609(15) & 123(42) &  \\
\hline
59362.2780166601(52) & 15.03 & 412.3(13) & 0.98 & 8.39(92) & 5.53(73)  & 9.9(11) &  &  &  \\
\hline
59362.2809834904(22) & 10.91 & 417.52(0) & 0.73 & 2.51(55) & 6.55(74)  & 5.12(71) &  &  &  \\
\hline
59362.28186508990 & 8.38 & 409.66(18) &  &  &  &  &  &   &  \\
\hline
59362.2828092863(19) & 29.67 & 419.09(39) & 1.97 & 7.04(32) & 2.064(15)  & 14.44(78) &  &  &  \\
\hline

\end{longtable}
}}

\vspace{-0.2cm}
{\footnotesize

Measured properties of the bursts detected from FRB\,20201124A during the uGMRT Band-4 observations. The table lists the burst arrival time (MJD), peak signal-to-noise ratio (S/N), power-optimised dispersion measure (DM), peak flux density ($S_{\rm peak}$), intrinsic width ($W_{\rm int}$), scattering timescale {($\tau_{\rm scat}$)}, fluence, peak emission frequency ($\nu_{\rm peak}$), bandwidth (BW), and drift rate. The burst arrival times were determined by fitting the frequency-integrated burst profiles with an exponentially convolved Gaussian model, using the burst time-of-arrival reported by the PRESTO single-pulse search as the initial estimate. The centroid of the best-fit model was adopted as the refined arrival time, and its uncertainty is quoted as the uncertainty in MJD. The peak S/N values are those reported by the PRESTO single-pulse search. Peak flux densities were estimated using the radiometer equation. The fluence was computed as $F = S_{\rm peak}W_{\rm eq}$, where the equivalent width is defined as $W_{\rm eq}=\sqrt{W_{\rm int}^{2}+ \tau_{\rm scat}^{2}}$. Bursts exhibiting multiple temporal components are listed in separate rows, with the intrinsic width measured independently for each component. Since the temporal separation between components is only a few milliseconds, a common scattering timescale was assumed for all components of a given burst. The scattering timescales were obtained from the frequency-averaged burst profiles over the 550--950 MHz band and are quoted at a reference frequency of 950 MHz. Blank entries indicate parameters that could not be reliably measured because of radio-frequency interference (RFI) or insufficient signal quality.}











\end{document}